\documentclass[aps,prb,twocolumn,showpacs,amssymb,amsfonts,amsmath,color,showpacs]{revtex4}

\usepackage{graphicx}
\usepackage{dcolumn}
\usepackage{bm}
\usepackage{color}

\begin{document}

\preprint{preprint(\today)}

\title{Study of the anisotropic magnetic order of the Eu sublattice in single crystals of EuFe$_{2-x}$Co$_{x}$As$_{2}$ ($x$=0, 0.2)  by means of magnetization and magnetic torque}

\author{Z.~Guguchia}
\email{zurabgug@physik.uzh.ch} \affiliation{Physik-Institut der
Universit\"{a}t Z\"{u}rich, Winterthurerstrasse 190, CH-8057
Z\"{u}rich, Switzerland}

\author{S.~Bosma}
\affiliation{Physik-Institut der Universit\"{a}t Z\"{u}rich,
Winterthurerstrasse 190, CH-8057 Z\"{u}rich, Switzerland}

\author{S.~Weyeneth}
\affiliation{Physik-Institut der Universit\"{a}t Z\"{u}rich,
Winterthurerstrasse 190, CH-8057 Z\"{u}rich, Switzerland}

\author{A.~Shengelaya}
\affiliation{Department of Physics, Tbilisi State University,
Chavchavadze 3, GE-0128 Tbilisi, Georgia}

\author{R.~Puzniak}
\affiliation{Institute of Physics, Polish Academy of Sciences,
Aleja Lotnik\'{o}w 32/46, PL-02-668 Warsaw, Poland}


\author{Z.~Bukowski}
\affiliation{Laboratory for Solid State Physics, ETH Z\"urich,
CH-8093 Z\"{u}rich, Switzerland}

\author{J.~Karpinski}
\affiliation{Laboratory for Solid State Physics, ETH Z\"urich,
CH-8093 Z\"{u}rich, Switzerland}

\author{H.~Keller}
\affiliation{Physik-Institut der Universit\"{a}t Z\"{u}rich,
Winterthurerstrasse 190, CH-8057 Z\"{u}rich, Switzerland }

\begin{abstract}
Here, we present a combination of magnetization and magnetic torque experiments to investigate the
magnetic orders in undoped EuFe$_2$As$_{2}$ and Co
doped EuFe$_{1.8}$Co$_{0.2}$As$_{2}$ single crystals. Although at
low temperatures typical results for an antiferromagnetic (AFM)
state in EuFe$_{2}$As$_{2}$ were found, our data strongly indicate
the occurrence of a canted antiferromagnetic (C-AFM) order of the
Eu$^{2+}$ moments between 17 K and 19 K, observed even in the lowest
studied magnetic fields.
However, unlike in the parent compound, no
low-field and low-temperature AFM state of the Eu$^{2+}$ moments was
observed in the doped EuFe$_{1.8}$Co$_{0.2}$As$_{2}$. Only a C-AFM
phase is present at low fields and low temperatures, with a reduced
magnetic anisotropy as compared to the undoped system. We present
and discuss for both, EuFe$_2$As$_{2}$ and
EuFe$_{1.8}$Co$_{0.2}$As$_{2}$, the experimentally deduced magnetic
phase diagrams of the magnetic ordering of the Eu$^{2+}$ sublattice
with respect to the temperature, the applied magnetic field, and its
orientation to the crystallographic axes. It is likely that the magnetic
coupling of the Eu and the Fe sublattice is strongly
depending on Co doping, having detrimental influence on the magnetic
phase diagrams as determined in this work. Their impact on the occurrence of superconductivity with
higher Co doping is discussed.

\end{abstract}


\maketitle

\section{Introduction}
 The discovery of superconductivity in
the iron-based pnictides \cite{Kamihara08} provided a new class of compounds to the
high temperature superconductor (HTS) family. Three main groups of these
iron-based superconductors are intensively studied: the $R$FeAsO compounds
with $R$ = La-Gd ('1111'),\cite{Kamihara08,Chen08} the ternary arsenides $A$Fe$_{2}$As$_{2}$ 
with $A$ = Ba, Sr, Ca, Eu ('122'),\cite{Rotter}
and the binary chalcogenides such as FeSe$_{1-x}$ ('11').\cite{Hsu} Similar to the cuprate HTS's,
the undoped iron-pnictides are not superconducting (SC) at ambient pressure and undergo a spin-density 
wave (SDW) transition at high temperatures.\cite{Xiao09} The SC state in iron-based compounds can be 
achieved either under pressure 
(chemical and hydrostatic) \cite{Torikachvili, Sun, Lee, Alireza, Igawa, Fukazawa, Duncan, Mani, Terashima, Miclea} 
or by appropriate charge carrier doping of the parent compounds,\cite{RenLu08,Matsuishi08,Zhao08}
both accompanied by a suppression of the SDW state.

 Here, we focus on EuFe$_{2}$As$_{2}$ which is a particularly interesting member of the ternary 
system $A$Fe$_{2}$As$_{2}$, since the $A$ site is occupied by a rare earth Eu$^{2+}$ $S$-state (orbital moment $L$ = 0) 
ion with a 4$f$$^{7}$ electronic configuration. Eu$^{2+}$ has a total
electron spin $S$ = 7/2, corresponding to a theoretical
effective magnetic moment of 7.94 ${\mu}$$_{\rm B}$. It is the only known member of the '122'
family containing 4$f$ electrons. In addition to the 
SDW ordering of the Fe moments at $T_{\rm SDW}$ ${\simeq}$ 190 K, an antiferromagnetic (AFM) 
order of the Eu$^{2+}$ spins 
at $T_{\rm AFM}$ ${\simeq}$ 19~K was reported by M\"{o}ssbauer and susceptibility measurements.\cite{Raffius93,ZRen,SJiang} 
Recently, neutron diffraction measurements 
were performed on EuFe$_{2}$As$_{2}$ and the magnetic structure illustrated in Fig.~1
was established.\cite{Xiao09} This material exhibits an \textit{A}-type 
AFM order of the Eu$^{2+}$ moments, $e.g.$, the Eu$^{2+}$ spins align
ferromagnetically in the planes, while the planes
are coupled antiferromagnetically.\cite{Xiao09,Blundell} 
It was demonstrated that by applying a high enough magnetic field,
the Eu$^{2+}$ moments can be realigned ferromagnetically
in both the parent compound EuFe$_{2}$As$_{2}$ \cite{SJiang,Xiao} as well as in the 
Co-doped system EuFe$_{2-x}$Co$_{x}$As$_{2}$ ($x$ = 0.22).\cite{ShuaiJiang} 
In addition, neutron diffraction measurements \cite{Xiao} suggested a canted AFM (C-AFM) 
structure of the Eu$^{2+}$ moments in EuFe$_{2}$As$_{2}$ at intermediate magnetic fields.

 Co-substitution may induce superconductivity in
EuFe$_{2-x}$Co$_{x}$As$_{2}$ with a reentrant behavior of resistivity
due to the AFM ordering of the Eu$^{2+}$ spins.\cite{He08}
Reentrant superconducting behavior was also
observed in resistivity experiments on EuFe$_{2}$As$_{2}$ under 
an applied pressure up to 2.5 GPa.\cite{Terashima,Miclea} However, 
only above 2.8 GPa, where a valence change of the Eu ions from a divalent magnetic 
state (4$f$$^{7}$, $J$ = 7/2) to a trivalent nonmagnetic state (4$f$$^{6}$, $J$ = 0) 
was suggested to occur,\cite{Sun} a sharp transition to a zero-resistivity state was observed.\cite{Terashima} 
Bulk superconductivity was also achieved in EuFe$_{2}$As$_{2-x}$P$_{x}$ \cite{Sun,Jeevan} where isovalent 
P-substitution of the As-site includes chemical pressure in EuFe$_{2}$As$_{2}$.
No superconductivity was detected in EuFe$_{2-x}$Ni$_{x}$As$_{2}$,\cite{ZRen09} while
superconductivity with a maximum $T_{\rm c}$ ${\simeq}$ 20 K was
reported for BaFe$_{2-x}$Ni$_{x}$As$_{2}$.\cite{LiLuo08} It was
suggested in various reports \cite{SJiang,Dengler10,Ying10,ZRen09} that 
there is a strong coupling between the localized Eu$^{2+}$
spins and the conduction electrons of the Fe$_{2}$As$_{2}$
layers. Recently, the hyperfine coupling constant $A_{\rm Eu}$ 
between the $^{75}$As nuclei and the Eu 4$f$ states in EuFe$_{1.9}$Co$_{0.1}$As$_{2}$ 
was quantitatively determined from $^{75}$As NMR to be
$A_{\rm Eu}$ = -1.9 ${\times}$ 10$^{7}$~A/m$\mu_{\rm B}$.\cite{Guguchia} 
This large value of $A_{\rm Eu}$ indicates
a strong coupling between the Eu$^{2+}$ localized moments and the charge 
carriers in the Fe$_{2}$As$_{2}$ layers   
and points to a strong correlation between the ordering of the localized 
magnetic moments and superconductivity in EuFe$_{2-x}$Co$_{x}$As$_{2}$.

 It is well established that the SDW state of the Fe moments is suppressed as a result   of Co doping. However,  
at present there is no clear picture how the ordering of the Eu spins develops with increasing
Co concentration. Generally, it was assumed 
that in the '122' systems the direction of the sublattice magnetization of the Eu$^{2+}$ 
magnetic moments is strongly affected by the magnetic behavior
of the Fe atoms.\cite{Xiao09,Nowik,NowikF,Feng,NowikFel,NowikFelner} 
Thus, it is important to compare the magnetic properties of the Eu-sublattice in     
EuFe$_{2-x}$Co$_{x}$As$_{2}$ without and with Co doping in order to study the 
correlation between ordering of Eu$^{2+}$ moments and the magnetism of the Fe sublattice. This in turn,
is crucial to understand the interplay between magnetism of localized moments and 
supercondcutivity in EuFe$_{2-x}$Co$_{x}$As$_{2}$.
 
 In this work, we present magnetic susceptibility, magnetization, and magnetic torque experiment performed on 
single crystals of EuFe$_{2-x}$Co$_{x}$As$_{2}$ ($x$ = 0, 0.2). 
The goal of this study is to investigate the macroscopic magnetic properties of the Eu-sublattice.
Magnetic susceptibility and magnetization investigations provide information on the magnetic structure
of a single-crystal sample in magnetic fields applied along the principal
axes. In addition, the evolution of the magnetic structure as a function
of the tilting angle of the magnetic field and the crystallographic axis can be studied by magnetic torque.
This paper is organized as follows: Experimental details are described in Sec.~II.
The results of the magnetic susceptibility, the magnetization and the magnetic torque 
measurements are presented and discussed in Sec.~III.
In Sec.~IV the magnetic phase diagrams of the Eu$^{2+}$ sublattice ordering with respect 
to magnetic field and temperature in single crystals of EuFe$_{2-x}$Co$_{x}$As$_{2}$ ($x$ = 0, 0.2) 
are discussed. The conclusions follow in Sec.~V.

\section{EXPERIMENTAL DETAILS}
 Single crystals of EuFe$_{2-x}$Co$_{x}$As$_{2}$ ($x$ = 0, 0.2) were grown out of Sn flux.\cite{Guguchia} 
The magnetization measurements
of the EuFe$_{2-x}$Co$_{x}$As$_{2}$ ($x$ = 0, 0.2) samples were performed with a commercial
SQUID magnetometer ($\textit{Quantum Design}$ MPMS-XL) with the magnetic field $H$ 
applied parallel ($H$ ${\parallel}$ $c$) or perpendicular ($H$ ${\perp}$ $c$) to the crystallographic $c$-axis. 
The magnetic torque measurements were carried out using a home-made torque sensor.\cite{Kohout2007} 
The sample is mounted on a platform hanging on piezoresistive legs. 
A magnetic field $\vec{H}$ applied to the sample
having magnetic moment $\vec{m}$ results in a mechanical 
torque $\vec{\tau}$ = $\mu_{0}$$\vec{m}$${\times}$$\vec{H}$.
This torque bends the legs, and thus creats a measurable electric signal proportional to the torque amplitude.
The temperature is controlled by an $\textit{Oxford}$ flow cryostat, and 
the magnetic field is provided by a rotatable resistive $\textit{Bruker}$ magnet 
with a maximum magnetic field of 1.4 T.
  

\section{RESULTS}
\subsection{Magnetization measurements}
\subsubsection{Temperature dependence}
\begin{figure}[t!]
\includegraphics[width=1\linewidth]{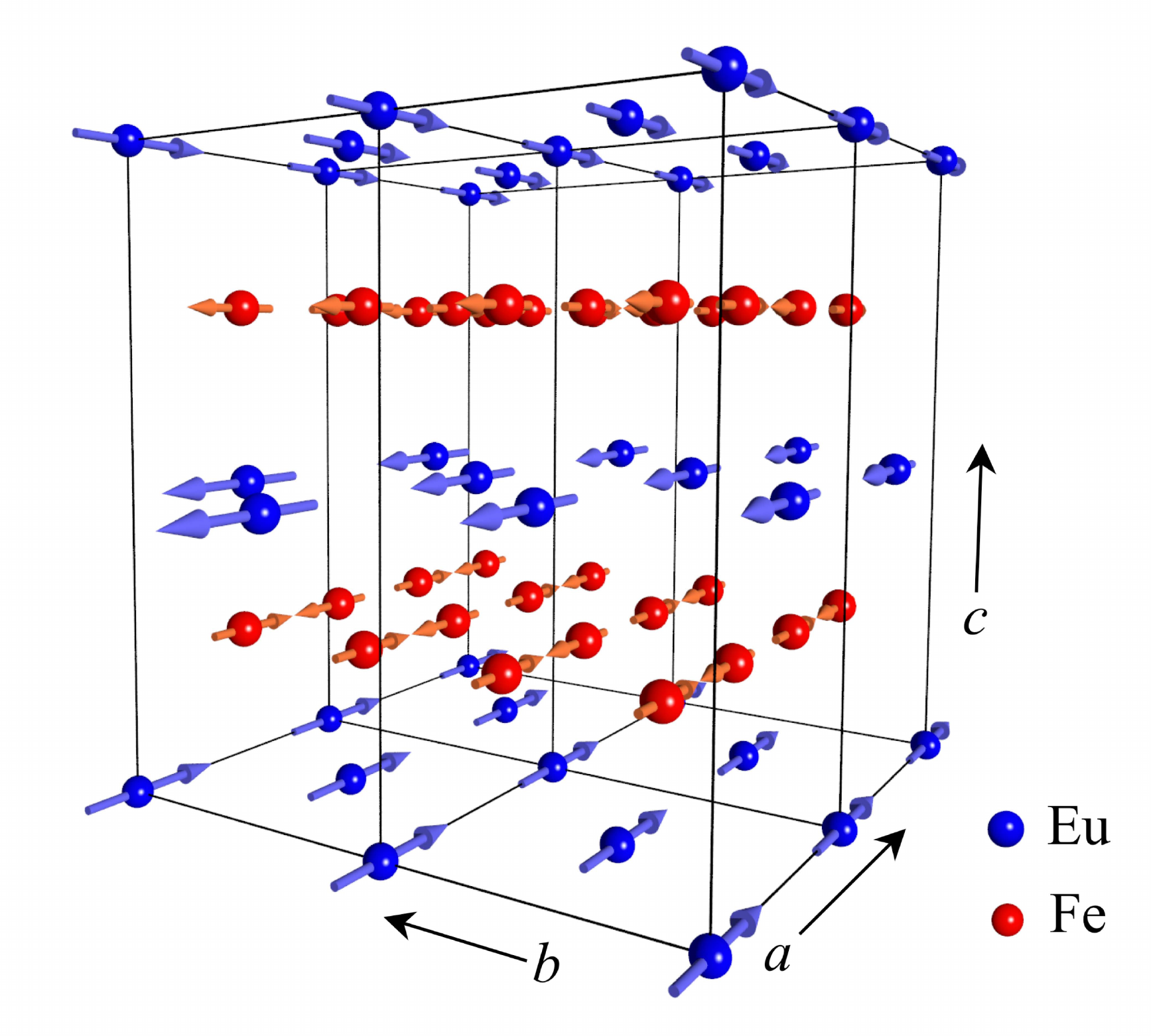}
\vspace{-0.2cm} \caption{ (Color online) Schematic illustration of the magnetic structure of EuFe$_{2}$As$_{2}$. 
The Fe moments (red) form a SDW state, whereas the Eu moments (blue) order ferromagnetically in the
$ab$-plane and align antiferromagnetically along the $c$-axis.} \label{fig2}
\end{figure}
\begin{figure}[t!]
\includegraphics[width=1\linewidth]{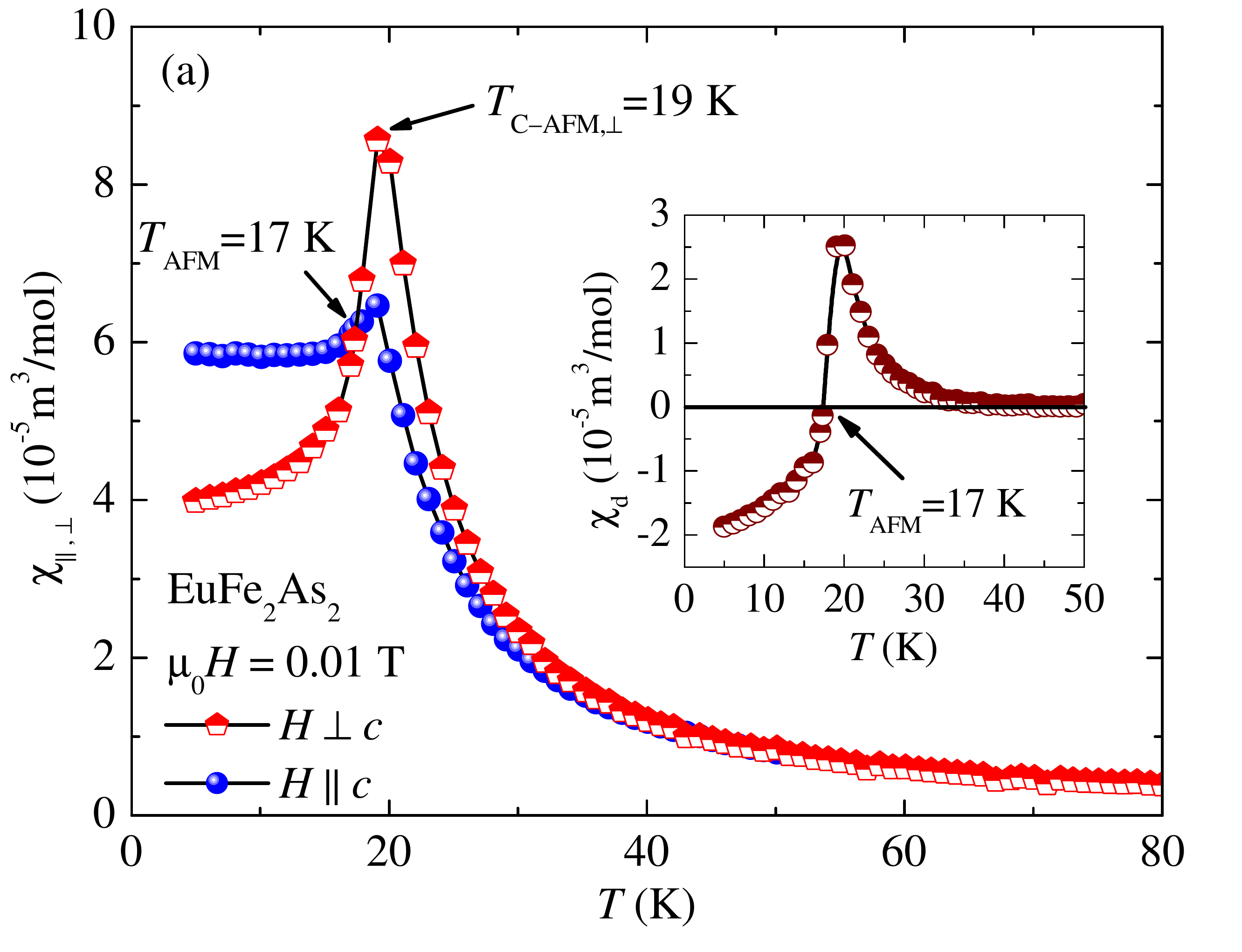}
\includegraphics[width=1\linewidth]{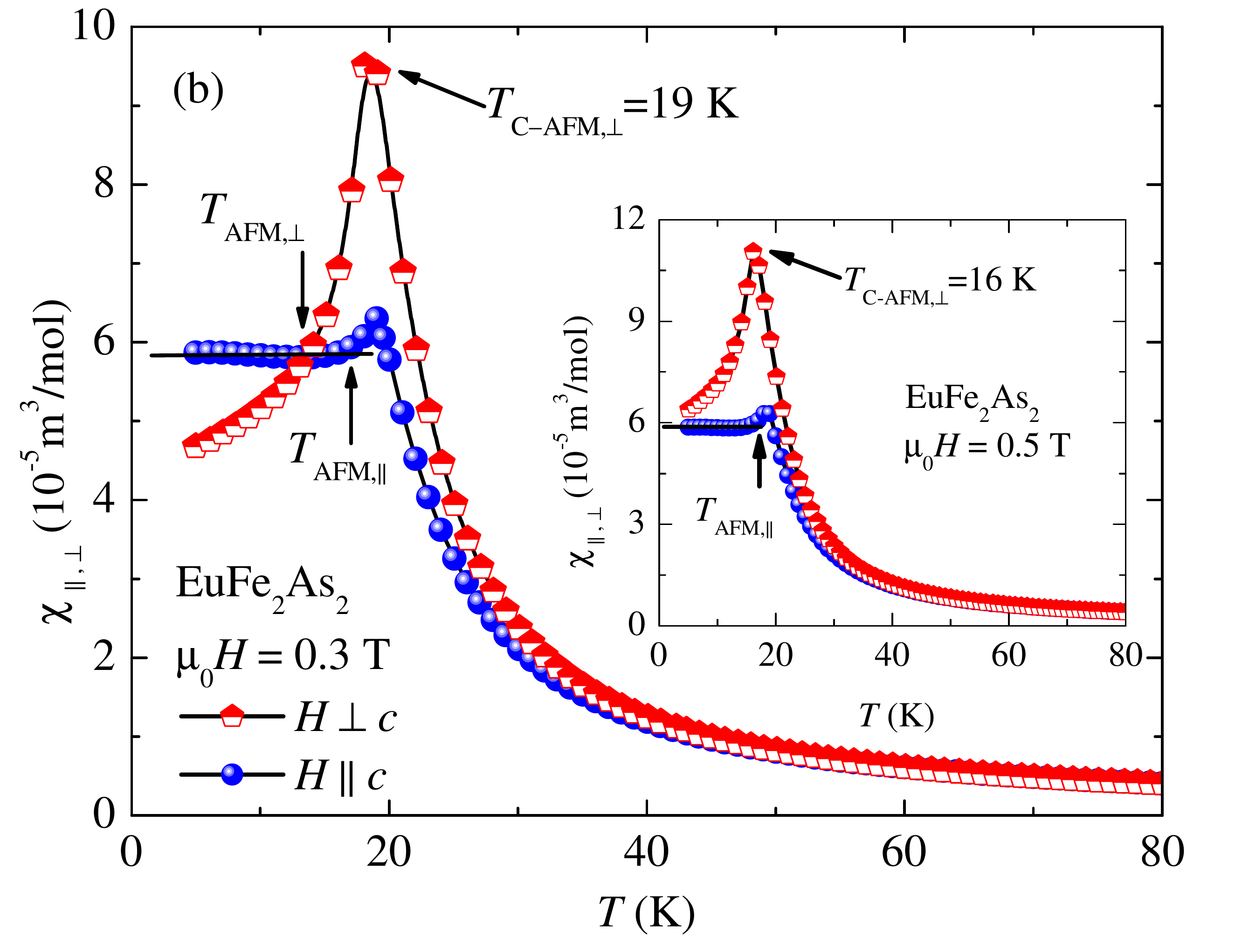}
\vspace{-0.5cm} \caption{ (Color online) Temperature dependence of the magnetic
susceptibility measured at fixed magnetic fields applied perpendicular (\textit{H} ${\perp}$ $\textit{c}$) and parallel (\textit{H} ${\parallel}$ $\textit{c}$) to the crystallographic $c$-axis of  
single-crystal EuFe$_{2}$As$_{2}$: (a) ${\mu}$$_{0}$$H$ = 0.01 T; (b) ${\mu}$$_{0}$$H$ = 0.3 T and ${\mu}$$_{0}$$H$ = 0.5 T (inset). 
The inset of panel (a) illustrates the temperature dependence of the difference between both susceptibilities (${\chi}_{\rm d} = {\chi}_{\perp} - {\chi}_{\parallel}$).     
The arrows mark the AFM and C-AFM ordering temperatures
of the Eu$^{2+}$ moments, and
$T_{\rm AFM}$,$_{\perp}$ and $T_{\rm AFM}$,$_{\parallel}$ refer to the AFM ordering temperatures for
\textit{H} ${\perp}$ $\textit{c}$ and \textit{H} ${\parallel}$ $\textit{c}$, respectively. 
The canted-AFM ordering temperature for 
\textit{H} ${\perp}$ $\textit{c}$ is denoted by $T_{\rm C-AFM}$,$_{\perp}$.} \label{fig2}
\end{figure}
\begin{figure}[t!]
\includegraphics[width=1\linewidth]{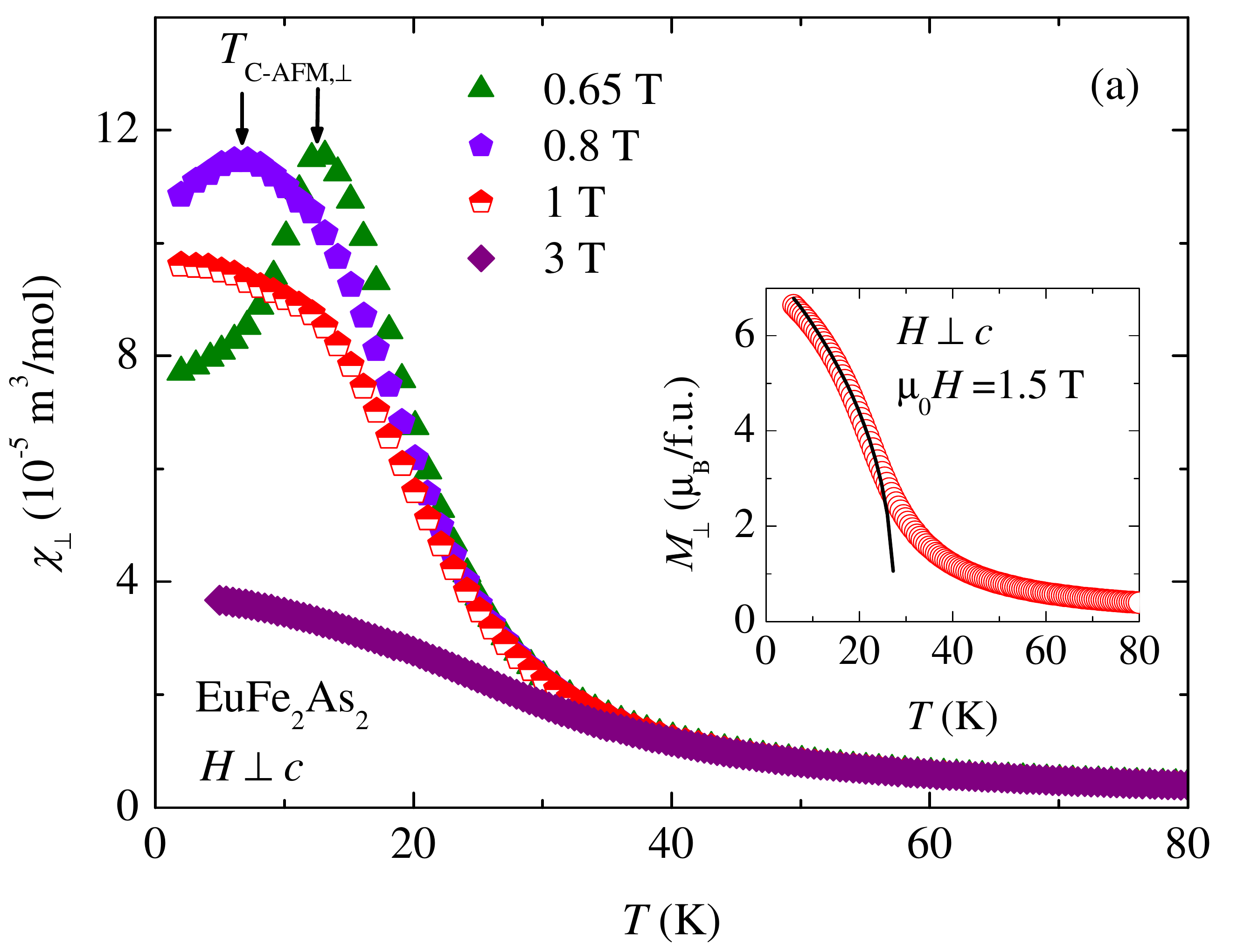}
\includegraphics[width=1\linewidth]{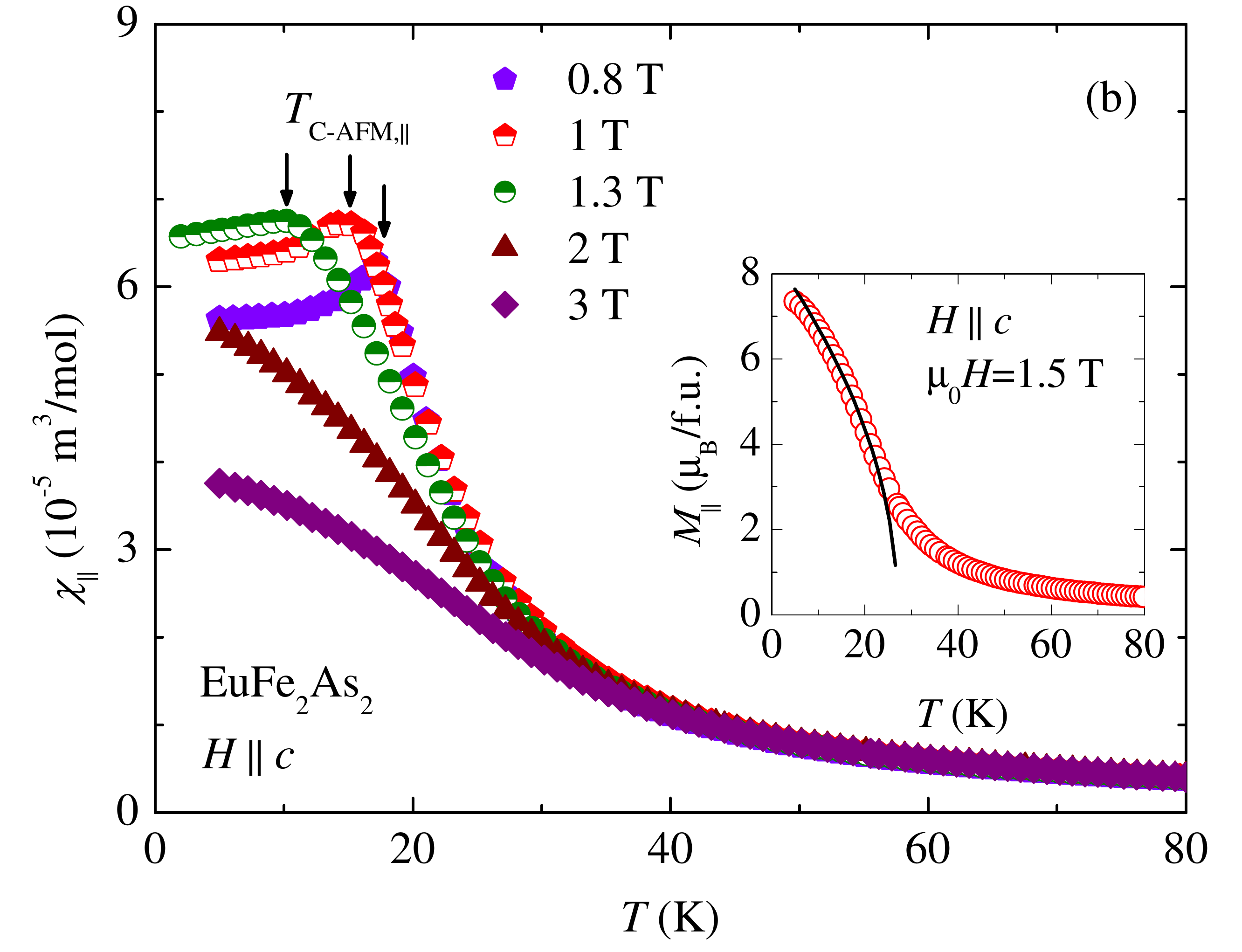}
\vspace{-0.3cm} \caption{ (Color online) Temperature dependence of the magnetic susceptibility measured at fixed magnetic fields of 
single-crystal EuFe$_{2}$As$_{2}$ for \textit{H} ${\perp}$ $\textit{c}$ (a) 
and \textit{H} ${\parallel}$ $\textit{c}$ (b). The arrows mark the canted antiferromagnetic ordering temperature $T_{\rm C-AFM}$ of the Eu$^{2+}$ moments in low fields. $T_{\rm C-AFM}$,$_{\perp}$ and $T_{\rm C-AFM}$,$_{\parallel}$ refer to the C-AFM ordering temperatures for
\textit{H} ${\perp}$ $\textit{c}$ and \textit{H} ${\parallel}$ $\textit{c}$, respectively. The insets illustrate the determination of $T_{\rm C}$ using the power law given in Eq.~(3).} \label{fig2}
\end{figure}
 The temperature dependence of the magnetic susceptibility ${\chi}$ = $M$/$H$ (here $M$ is the magnetization 
determined as magnetic moment per mol) for the crystal  of EuFe$_{2}$As$_{2}$ in a field of ${\mu}$$_{0}$\textit{H} = 0.01 T for \textit{H} ${\perp}$ \textit{c} and 
for \textit{H} ${\parallel}$ \textit{c} is shown in Fig.~2a. 
In agreement with previous reports,\cite{ZRen,SJiang} the magnetic susceptibility for \textit{H} ${\perp}$ $\textit{c}$ (${\chi}$$_{\perp}$) and for  \textit{H} ${\parallel}$ $\textit{c}$ (${\chi}$$_{\parallel}$), determined in the temperature
range from 30 to 190 K ($i.e.$, far above $T_{\rm AFM}$ ${\simeq}$ 19 K of the Eu moments 
up to $T_{\rm SDW}$ ${\simeq}$ 190 K of the Fe moments) 
is well described by the Curie-Weiss law:
\begin{equation}
\chi_{\rm}(T)=\frac{C}{T-\theta_{\rm CW}} \label{eq1}.
\end{equation}
 Here, ${\textit C}$ denotes the Curie constant and ${\theta}$$_{\rm CW}$ the Curie-Weiss temperature. 
Analyzing the data in Fig.~2a with Eq.~(1) yields:
\textit{C} = 1853(15)${\times}$10$^{-7}$ m$^{3}$ K/mol, ${\theta}$$_{\rm CW}$ = 19.74(8) K for \textit{H} ${\parallel}$ $\textit{c}$
and \textit{C} = 2127(23)${\times}$10$^{-7}$ m$^{3}$ K/mol, ${\theta}$$_{\rm CW}$ = 20.69(4) K for \textit{H} ${\perp}$ $\textit{c}$. The calculated effective magnetic moment is
${\mu}$$_{\rm eff}$ ${\simeq}$ 7.6 ${\mu}$$_{\rm B}$ for \textit{H} ${\parallel}$ $\textit{c}$ and ${\mu}$$_{\rm eff}$ ${\simeq}$ 8.3 ${\mu}$$_{\rm B}$ for \textit{H} ${\perp}$ $\textit{c}$. These estimates of ${\mu}$$_{\rm eff}$ are
close to the theoretical value of the magnetic moment of a free Eu$^{2+}$ ion 
(${\mu}$$_{\rm Eu^{2+}}$ = 7.94~${\mu}$$_{\rm B}$). 
The positive value of ${\theta}$$_{\rm CW}$ for both 
\textit{H} ${\parallel}$ $\textit{c}$ and 
\textit{H} ${\perp}$ $\textit{c}$ is consistent with previous magnetization 
measurements,\cite{ZRen,SJiang} indicating that the direct interaction between 
the Eu$^{2+}$ moments is ferromagnetic (FM). This is in agreement with the magnetic 
structure of EuFe$_{2}$As$_{2}$ suggested by zero field neutron diffraction measurements,\cite{Xiao09} 
revealing that the intralayer arrangment of the Eu$^{2+}$ spins is FM. 
The sharp increase of ${\chi}$ with decreasing temperature below 30 K also 
indicates a FM coupling between the Eu$^{2+}$ moments. 
The Eu moments align with respect to the Fe moments along the 
a axis \cite{Xiao09} as illustrated in Fig.~1.

 With decreasing temperature from 19~K to 17~K, the susceptibility ${\chi}$$_{\perp}$ of
single-crystal EuFe$_{2}$As$_{2}$ decreases rapidly and below 17~K the decrease 
of ${\chi}$$_{\perp}$ is less pronounced. On the other hand, ${\chi}$$_{\parallel}$ decreases with decreasing temperature from 19~K to 17~K and remains constant below 17~K.
Moreover, the values of ${\chi}$$_{\perp}$ and ${\chi}$$_{\parallel}$ at 19 K are substantially different (${\chi}$$_{\perp}$/${\chi}$$_{\parallel}$ ${\simeq}$ 1.33), already in a rather low magnetic field ${\mu}$$_{0}$$H$ = 0.01 T (see Fig.~2a). Note that within the classical picture of an ideal antiferromagnet, the magnetic susceptibility ${\chi}$ in a magnetic field perpendicular to the easy axis is constant, and ${\chi}$ in a field parallel to the easy plane decreases linearly with decreasing temperature. In addition, 
for an antiferromagnet the values of ${\chi}$ at the antiferromagnetic (AFM) transition temperature are
the same for both \textit{H} ${\perp}$ $\textit{c}$ and \textit{H} ${\parallel}$ $\textit{c}$.\cite{Blundell}
The inset of Fig.~2a illustrates the temperature dependence of the difference between both susceptibilities ${\chi}_{\rm d} = {\chi}_{\perp} - {\chi}_{\parallel}$.
Note that below 19~K the quantity 
${\chi}$$_{\rm d}$ decreases with decreasing temperature
and reaches zero at around 17 K.
This behavior of ${\chi}$$_{\rm d}$($T$) can be explained by invoking a transition from the high-temperature paramagnetic state to
a FM state or to a C-AFM state at about 19 K. The transition from a FM or a C-AFM to an AFM state of the Eu$^{2+}$ spins occurs only below 17 K. 
The pronounced increase of ${\chi}$$_{\parallel}$ above 17 K indicates the appearence of a magnetic moment along the $c$-axis. 
Since ${\chi}$$_{\parallel}$ is smaller than ${\chi}$$_{\perp}$ in the FM/C-AFM state, 
it is suggested that the $a$$b$-plane is the easy plane of this ordered state. 
In Fig.~2b the temperature dependences of ${\chi}$$_{\perp}$ and ${\chi}$$_{\parallel}$ of single-crystal
EuFe$_{2}$As$_{2}$ in a magnetic field of 0.3 T and 0.5 T (inset) are shown. Obviously, the AFM transition temperatures for \textit{H} ${\perp}$ \textit{c}
(crossing point of ${\chi}$$_{\perp}$ and ${\chi}$$_{\parallel}$) and for \textit{H} ${\parallel}$ \textit{c} (temperature at which ${\chi}$$_{\parallel}$ starts to increase) are shifted to lower temperature with
higher magnetic field (see Fig.~2a for comparison). However, at ${\mu}$$_{0}$$H$ = 0.5 T the curves
${\chi}$$_{\perp}$ and ${\chi}$$_{\parallel}$ do not cross in the investigated temperature range,
indicating that the AFM state of the Eu$^{2+}$ ions 
is suppressed in EuFe$_{2}$As$_{2}$ in magnetic fields \textit{H} ${\perp}$ \textit{c} exceeding ${\mu}$$_{0}$$H$ ${\simeq}$ 0.5 T. For \textit{H} ${\parallel}$ \textit{c} the suppresion of the AFM state occurs in fields higher than ${\mu}$$_{0}$$H$ ${\simeq}$ 1.2 T, since above this field the susceptibility for
\textit{H}~${\parallel}$~\textit{c} is temperature dependent even at temperature as low
as 2 K (see Fig.~3b). Importantly, the magnetic field at which the magnetic moments of the Eu sublattice saturates
($i.e.$, the field at which the FM state is reached) is much higher than the field
of suppresion of the AFM state. 
This implies that a FM state appears in a magnetic field higher than the field of suppression of antiferromagnetism
and that those two transitions are distinguishable. The peak in the magnetic susceptibility 
at about 19 K in low fields (see Fig.~2) can be associated with the transition from a PM to a C-AFM state.
This peak is shifted to lower temperature with applied 
magnetic field above ${\mu}$$_{0}$$H$ ${\simeq}$ 0.3 T for \textit{H} ${\perp}$ \textit{c} and above ${\mu}$$_{0}$$H$ ${\simeq}$ 0.5 T for \textit{H} ${\parallel}$ \textit{c} (see Figs.~2b and 3b).
Finally, we may conclude
that a field-induced magnetic phase transition from an AFM via a C-AFM configuration to a FM state takes place
below 17 K. Such a transition is visible even at the lowest temperature of 2 K reached in our experiment.

 The magnetization $M$($T$) in the FM state in the vicinity of the Curie temperature $T_{\rm C}$ 
can be described by the power law:
\begin{equation}
M(T)=M_0\left(1-\frac{T}{T_{\rm C}}\right)^{\tilde{\beta}} \label{eq1}.
\end{equation}
 Here ${\tilde{\beta}}$ and $M_{0}$ are empirical constants. 
Analyzing the data at 1.5 T with Eq.(2) yields: $T_{\rm C}$ = 27.2(1) K and ${\tilde{\beta}}$ = 0.39(1) for both directions of the magnetic field (solid lines in the insets of Fig.~3a and 3b). 
It was found that $T_{\rm C}$ increases gradually with increasing applied magnetic field for \textit{H} ${\perp}$ $\textit{c}$ and \textit{H} ${\parallel}$ $\textit{c}$. By extrapolating $T_{\rm C}$($H$) to low fields 
the zero-field value of $T_{\rm C}$ was found to be ${\simeq}$ 19 K. The present values of $T_{\rm C}$($H$) are in agreement with those reported by Xiao $et$ $al$.\cite{Xiao} 

\begin{figure}[b!]
\includegraphics[width=1\linewidth]{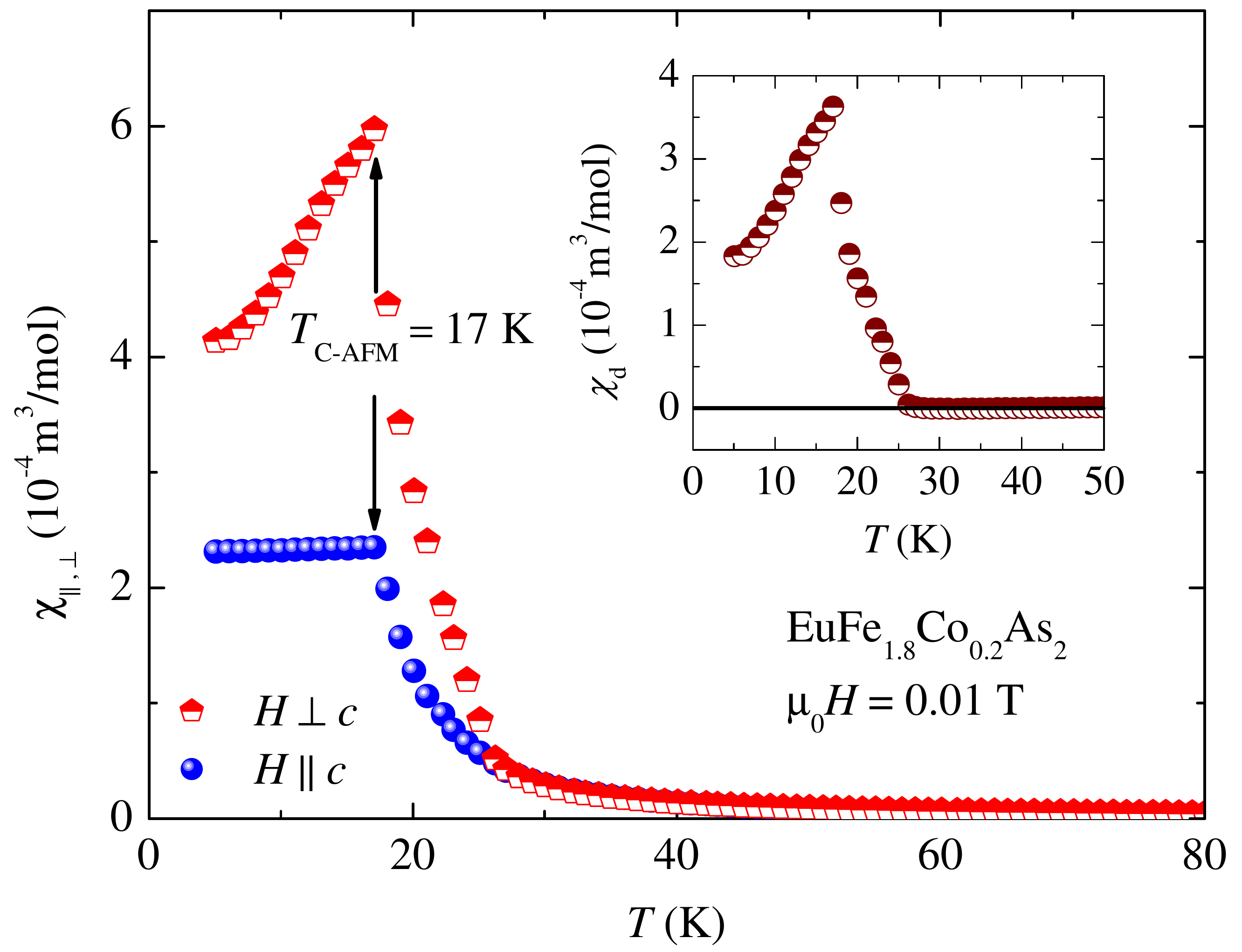}
\vspace{-0.2cm} \caption{ (Color online) Temperature dependence of the magnetic susceptibility measured in
a field of ${\mu}$$_{0}$$H$ = 0.01 T of single-crystal EuFe$_{1.8}$Co$_{0.2}$As$_{2}$ for  \textit{H} ${\perp}$ $\textit{c}$ and \textit{H} ${\parallel}$ $\textit{c}$. 
The arrows mark the canted antiferromagnetic ordering temperature $T_{\rm C-AFM}$ ${\simeq}$ 17 K of
the Eu$^{2+}$ moments. In the inset the difference between the susceptibilities for the 
two different field configurations (${\chi}_{\rm d} = {\chi}_{\perp} - {\chi}_{\parallel}$)
is plotted as a function of temperature.} \label{fig7}
\end{figure}
\begin{figure}[t!]
\includegraphics[width=1\linewidth]{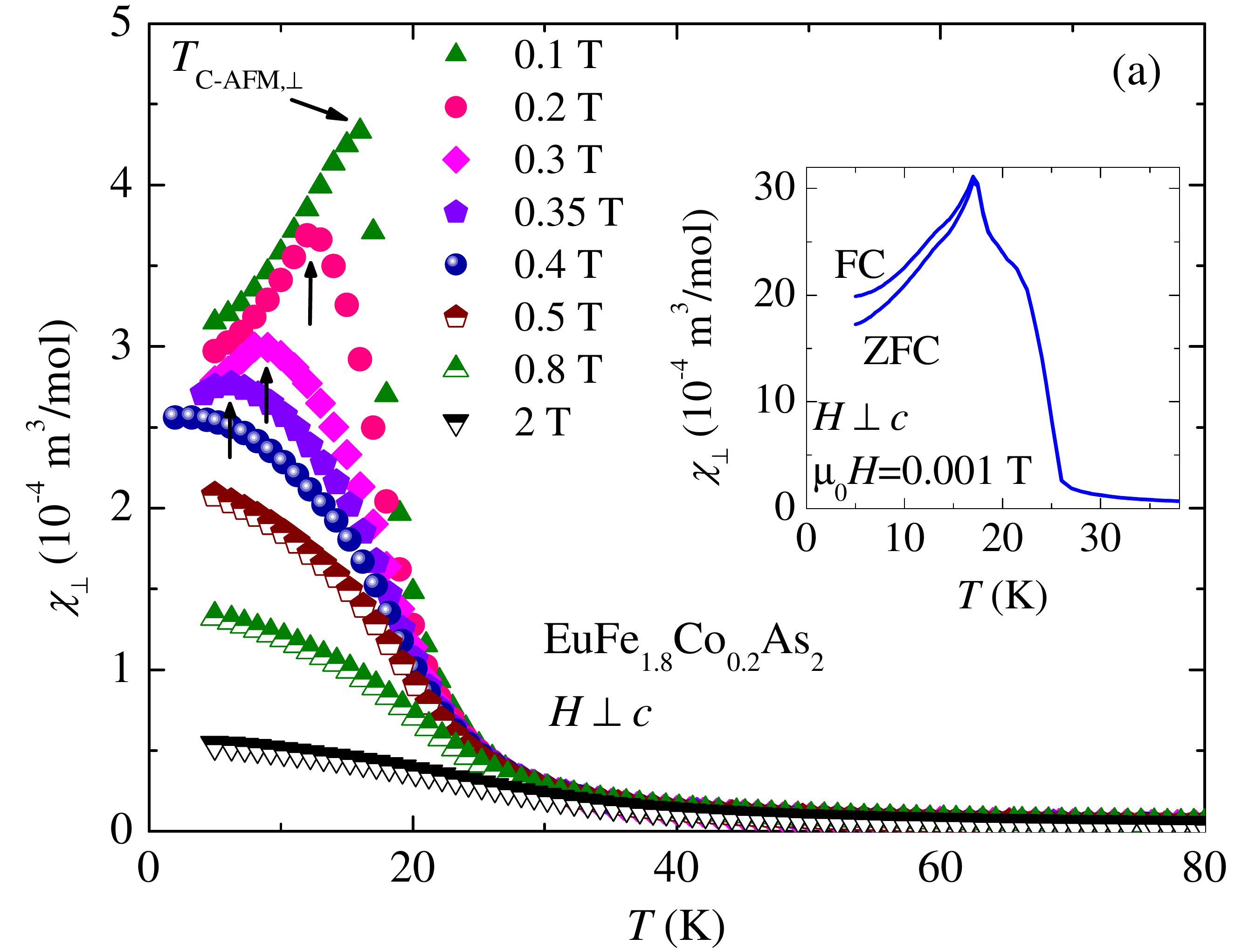}
\includegraphics[width=1\linewidth]{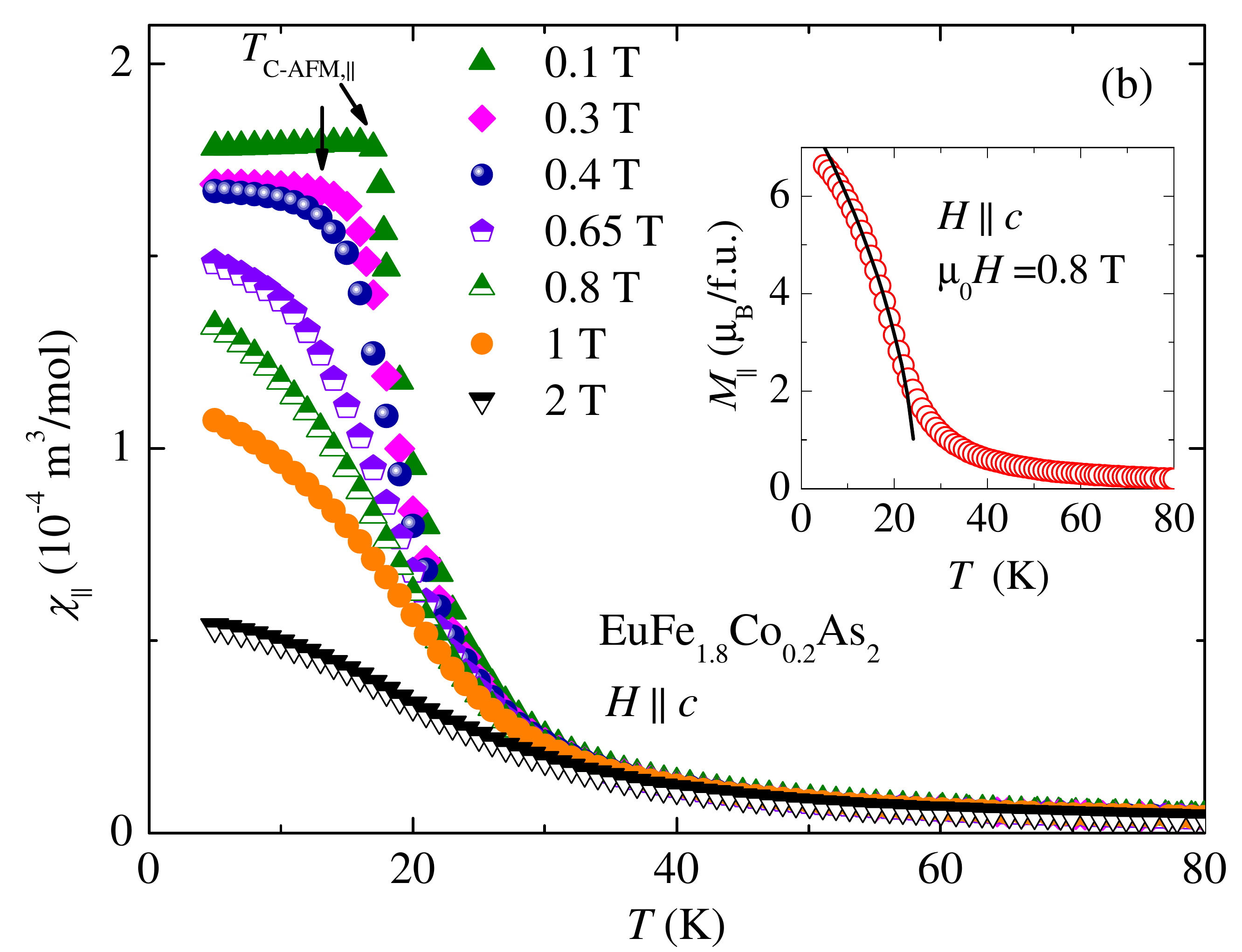}
\vspace{-0.2cm} \caption{ (Color online) Temperature dependence of the ZFC magnetic susceptibility measured at various fixed magnetic fields of single-crystal
EuFe$_{1.8}$Co$_{0.2}$As$_{2}$ for  \textit{H} ${\perp}$ $\textit{c}$ (a)
and \textit{H} ${\parallel}$ $\textit{c}$ (b). 
The arrows mark the canted antiferromagnetic ordering temperature $T_{\rm C-AFM}$ of
the Eu$^{2+}$ moments in low magnetic fields.
$T_{\rm C-AFM}$,$_{\perp}$ and $T_{\rm C-AFM}$,$_{\parallel}$ refer to the C-AFM ordering temperatures for
\textit{H} ${\perp}$ $\textit{c}$ and \textit{H} ${\parallel}$ $\textit{c}$, respectively. In the inset of (a) ${\chi}$$_{\rm \perp}$($T$) for FC and ZFC
in an applied field of ${\mu}$$_{0}$$H$ = 0.001 T is plotted. The inset of (b) shows the approximation of 
$M$$_{\parallel}$($T$) in ${\mu}$$_{0}$$H$ = 0.8 T by the power law (solid curve) given in Eq.~(3).} \label{fig7}
\end{figure}
  The temperature dependence of the magnetic susceptibility for the Co doped crystal EuFe$_{1.8}$Co$_{0.2}$As$_{2}$ in an applied field of ${\mu}$$_{0}$$H$ = 0.01 T for 
\textit{H} ${\perp}$ \textit{c} and \textit{H} ${\parallel}$ \textit{c} is presented in Fig.~4. 
In the inset the temperature dependence of the difference between the susceptibilities for two
field configurations ${\chi}_{\rm d} = {\chi}_{\perp} - {\chi}_{\parallel}$ is shown. 
Analyzing the susceptibility data above 30 K with Eq.~(2) yields: 
\textit{C} = 2108(32)${\times}$10$^{-7}$ m$^{3}$ K/mol, ${\theta}$$_{\rm CW}$ = 21.86(6) K for 
\textit{H} ${\perp}$ $\textit{c}$ and \textit{C} = 1915(34)${\times}$10$^{-7}$ m$^{3}$ K/mol, 
${\theta}$$_{\rm CW}$ = 20.67(7) K for 
\textit{H}~${\parallel}$ $\textit{c}$. Again ${\theta}$$_{\rm CW}$ turns out to be
positive. Like in the parent compound a sharp increase of ${\chi}$ below 30 K is observed,
which is attributed to the in-plane FM coupling between the Eu$^{2+}$ moments. Below 17 K the
susceptibility ${\chi}$$_{\perp}$ starts to decrease with decreasing temperature, indicating
the onset of an AFM transition of the Eu$^{2+}$ spins. On the other hand, ${\chi}$$_{\parallel}$ 
remains almost constant below 17 K. This suggests that the Eu$^{2+}$ moments align along the $a$$b$-plane, 
similar to undoped EuFe$_{2}$As$_{2}$. However, for EuFe$_{2}$As$_{2}$ the AFM ordering temperature 
of the Eu$^{2+}$ spins is about 2 K higher.
\begin{figure}[b!]
\includegraphics[width=1\linewidth]{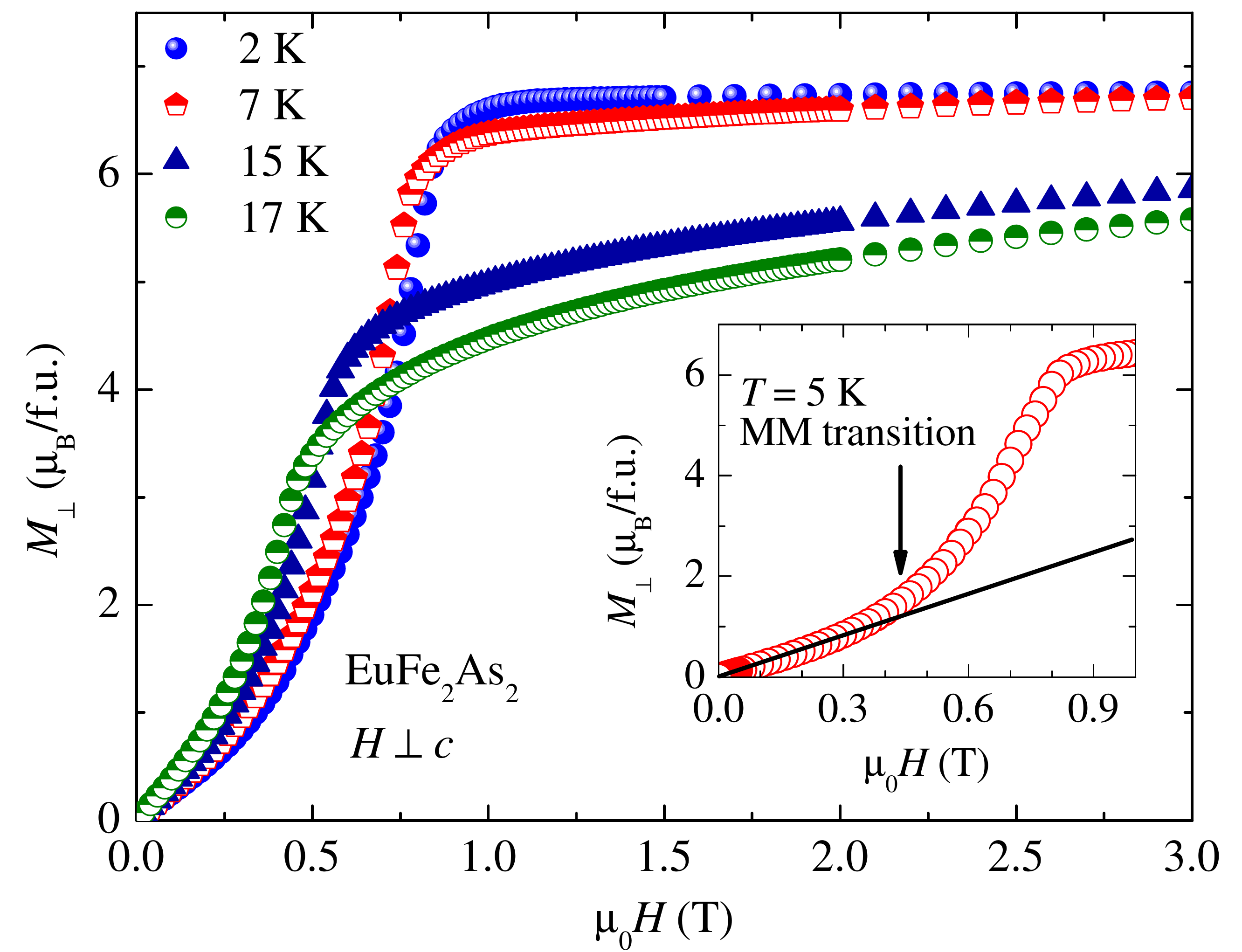}
\vspace{-0.2cm} \caption{ (Color online) Field dependence of the magnetization at various temperatures of 
single-crystal EuFe$_{2}$As$_{2}$ for \textit{H} ${\perp}$ $\textit{c}$. The inset shows the low field 
$M_{\rm \perp}$ data at 5 K, illustrating the metamagnetic (MM) transition marked by the arrow.} \label{fig2}
\end{figure}
\begin{figure}[b!]
\includegraphics[width=1\linewidth]{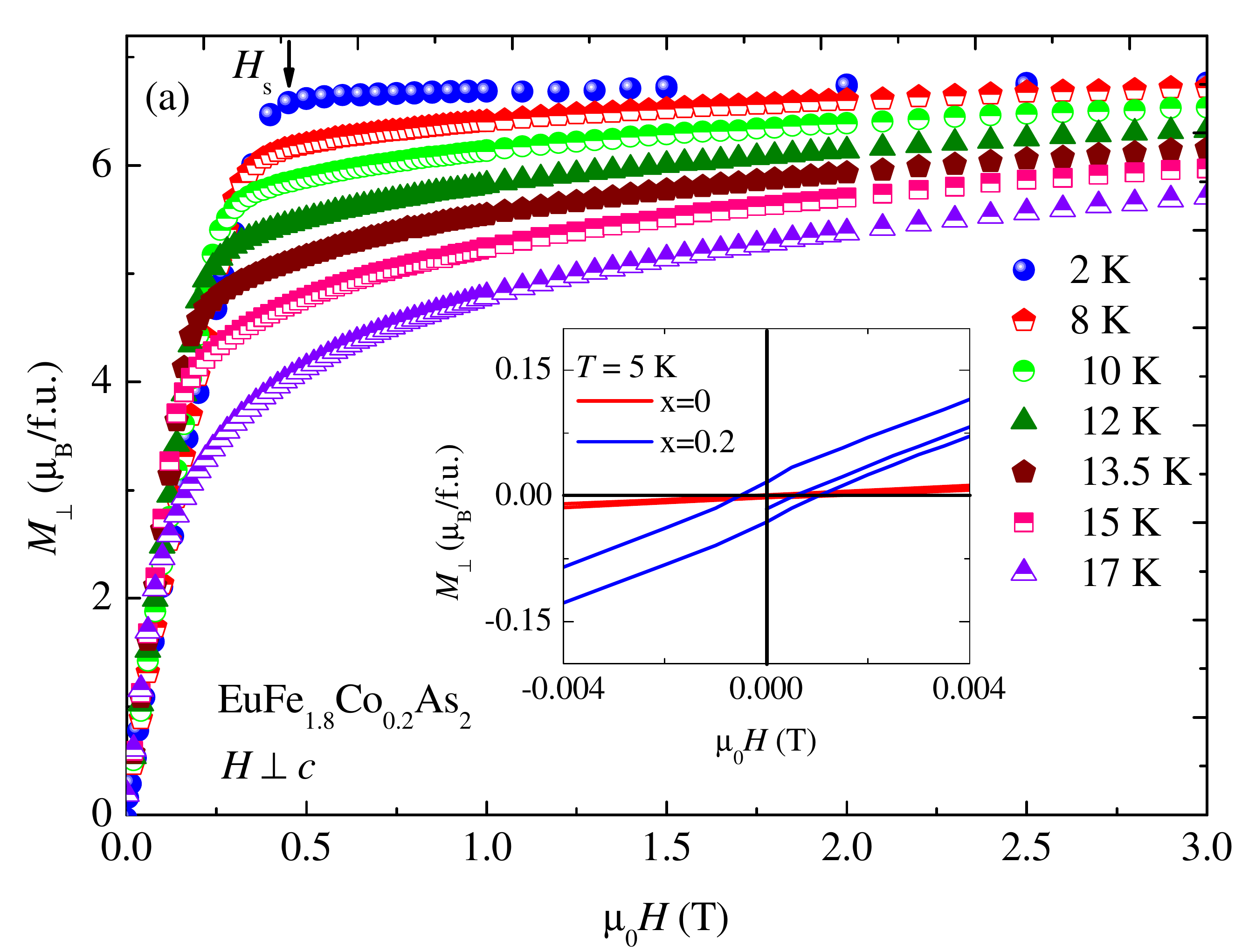}
\includegraphics[width=1\linewidth]{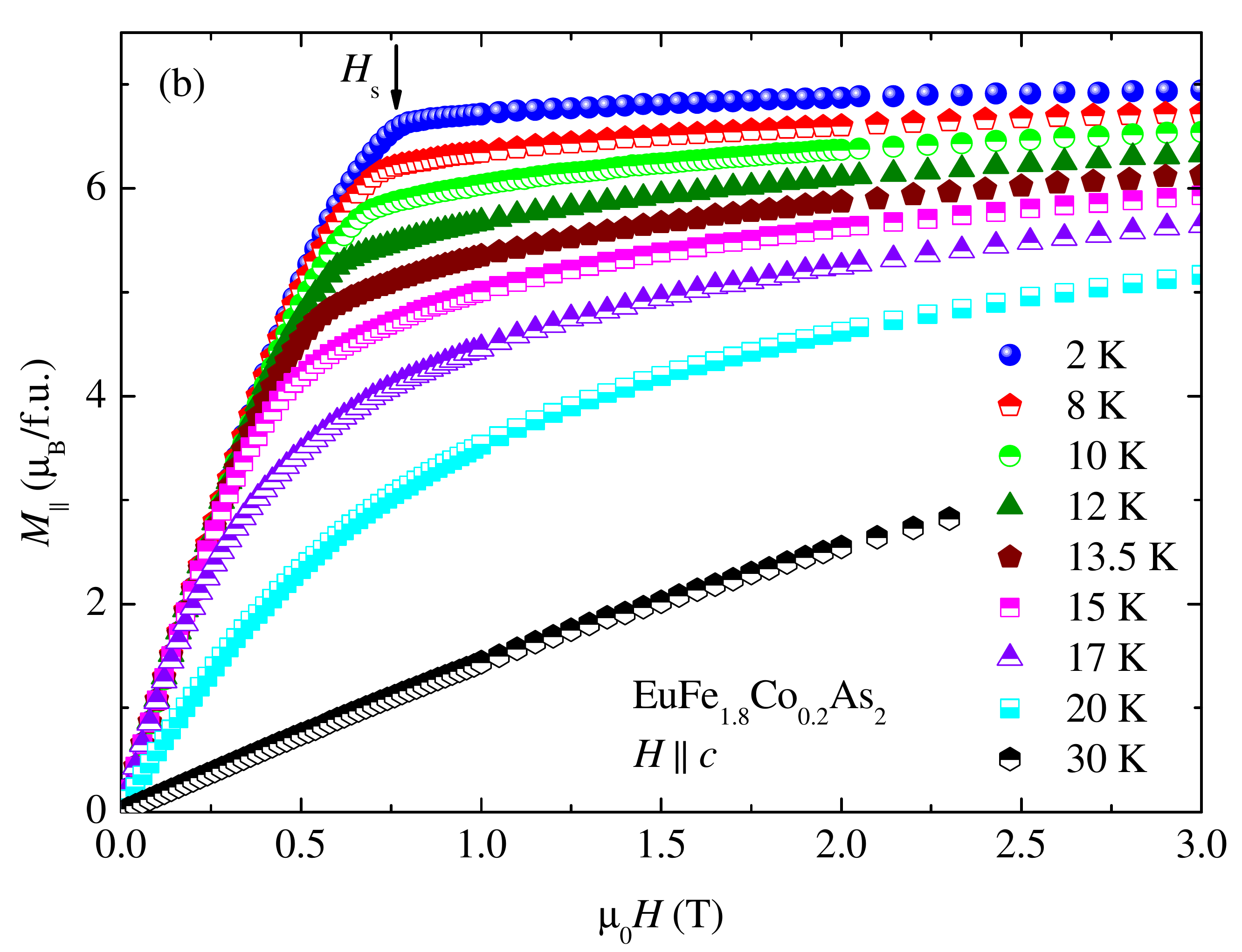}
\vspace{-0.2cm} \caption{ (Color online) Field dependence of the magnetization at low 
temperatures of single-crystal EuFe$_{1.8}$Co$_{0.2}$As$_{2}$ for
\textit{H} ${\perp}$ $\textit{c}$ (a) and \textit{H} ${\parallel}$ $\textit{c}$ (b). 
The saturation field $H_{\rm s}$ at 2 K is marked by arrows. The inset of (a) shows 
the field dependence of $M_{\rm \perp}$ for EuFe$_{2}$As$_{2}$ and EuFe$_{1.8}$Co$_{0.2}$As$_{2}$ at 5 K.} \label{fig2}
\end{figure}
 Below 17 K, ${\chi}$$_{\perp}$ is significantly larger than ${\chi}$$_{\parallel}$, even in magnetic fields as low as ${\mu}$$_{0}$${\textit{H}}$ = 0.01 T (see Fig.~4). 
Thus, no crossing between ${\chi}$$_{\perp}$ and
${\chi}$$_{\parallel}$ is observed (inset of Fig.~4), in contrast to the parent compound EuFe$_{2}$As$_{2}$ 
(see Fig.~2). Furthermore, ${\chi}$$_{\perp}$ is temperature dependent even at the lowest applied magnetic field. 
This is inconsistent with an AFM state with an easy $c$-axis. Hence, we suggest that for all temperatures below 17 K the ground state of the coupled Eu$^{2+}$ spins in EuFe$_{1.8}$Co$_{0.2}$As$_{2}$ is a C-AFM state with a FM component in the $a$$b$-plane. 
This implies that the magnetic configuration of the Eu moments is strongly influenced by the
magnetization of the Fe-sublattice. This is consistent with previous NMR studies, revealing a 
strong coupling between the Eu and {Fe$_{2-x}$Co$_{x}$As$_{2}$} layers.\cite{Guguchia} 

 The temperature dependences of ${\chi}$$_{\perp}$ and ${\chi}$$_{\parallel}$ at different 
magnetic fields of single-crystal EuFe$_{1.8}$Co$_{0.2}$As$_{2}$ are shown in Fig.~5. 
Zero field cooling (ZFC) and field cooling (FC) susceptibilities ${\chi}$$_{\perp}$($T$) measured
in an applied field of ${\mu}$$_{0}$$H$ = 0.001 T are shown in the inset of Fig.~5a. 
Below 17 K the ZFC and FC curves deviate from each other, indicating
the presence of a C-AFM state of the Eu$^{2+}$ moments. 
The data reveal a decrease of the C-AFM ordering temperature $T_{\rm C-AFM}$ with increasing magnetic field for both field orientations, similar as for the parent compound EuFe$_{2}$As$_{2}$. However, the values
for $T_{\rm C-AFM}$ for EuFe$_{1.8}$Co$_{0.2}$As$_{2}$ are substantially smaller than those for EuFe$_{2}$As$_{2}$.

\subsubsection{Field dependence}

 The susceptibility investigations of the previous section clearly demonstrate that the system  
EuFe$_{2-x}$Co$_{x}$As$_{2}$ ($x$ = 0, 0.2)
shows a rich variety of magnetic phases. In order to explore in detail the various magnetic field-induced
phases, magnetization experiments were also performed as a function of the applied magnetic field at different temperatures. 

 The field dependence of the magnetization of single-crystal EuFe$_{2}$As$_{2}$ at different temperatures for 
\textit{H}~${\perp}$ $\textit{c}$ is shown in Fig.~6. In the inset the low field magnetization
$M_{\rm \perp}$ at 5 K is shown. $M_{\rm \perp}$ increases almost linearly with increasing magnetic field 
$H$ up to ${\mu}$$_{0}$$H$ ${\simeq}$ 0.45 T where a sudden increase of $M_{\rm \perp}$ appears. 
Then $M_{\rm \perp}$ further increases with increasing $H$, and finally
saturates for ${\mu}$$_{0}$$H$ ${\geq}$ 0.8 T. 
The value of the saturation magnetization corresponds to an effective magnetic moment of 6.8 ${\mu}$$_{\rm B}$/f.u., 
which is close to g${\mu}$$_{\rm B}$$S$ = 7 ${\mu}$$_{\rm B}$/f.u. expected for Eu$^{2+}$ moments.
This result suggests that there is a metamagnetic (MM) \cite{Luo,Perry} transition at ${\mu}$$_{0}$$H_{\rm MM}$ ${\simeq}$ 0.45 T at 5 K in EuFe$_{2}$As$_{2}$, consistent with previous observations.\cite{ZRen,SJiang} Such a metamagnetic transition is characteristic for 
A-type antiferromagnetism in layered systems as, $e.g.$, 
Na$_{0.85}$CoO$_{2}$ \cite{Luo} and La$_{2-x}$Sr$_{1+x}$Mn$_{2}$O$_{7}$.\cite{Kimura} Figure~6 shows that the MM transition shifts towards
lower fields with increasing temperature. The values of the magnetic field at which the MM transition
occurs is in agreement with the results obtained
from the susceptibility for the AFM to C-AFM transition.
Thus, we propose that the MM transition corresponds to the onset of a spin-flop
transition \cite{Blundell} from an AFM to a C-AFM state in EuFe$_{2}$As$_{2}$.
However, no MM transition for \textit{H} ${\perp}$ $\textit{c}$ is detected in EuFe$_{1.8}$Co$_{0.2}$As$_{2}$ (Fig.~7a). 
Both $M_{\rm \perp}$ and $M_{\rm \parallel}$ first increase almost linearly with increasing $H$ and then saturate  at higher fields (Fig.~7).
The absence of a MM transition in EuFe$_{1.8}$Co$_{0.2}$As$_{2}$ is consistent with the susceptibility measurements presented above,
suggesting that the Eu$^{2+}$ moments exhibit a C-AFM ground state even at very low $H$.
This conclusion is also supported by magnetic hysteresis measurements at 5 K performed in magnetic
fields up to 0.5 T. As demonstrated in the inset of Fig.~7a the field dependence of 
$M_{\rm \perp}$ at 5 K shows a well developed hysteresis for EuFe$_{1.8}$Co$_{0.2}$As$_{2}$, in contrast to the
parent compound EuFe$_{2}$As$_{2}$ where no hysteresis is observed.

 Obviously, the presented susceptibility and magnetization measurements reveal
a complex and rather sophisticated interplay of magnetic phases in the EuFe$_{2-x}$Co$_{x}$As$_{2}$ system.
Additional information on the complex magnetic phases in EuFe$_{2-x}$Co$_{x}$As$_{2}$ is
obtained from angular dependent magnetic torque studies presented in the next section.

\subsection{Magnetic torque}

\begin{figure}[b]
\centering
\includegraphics[width=1\linewidth]{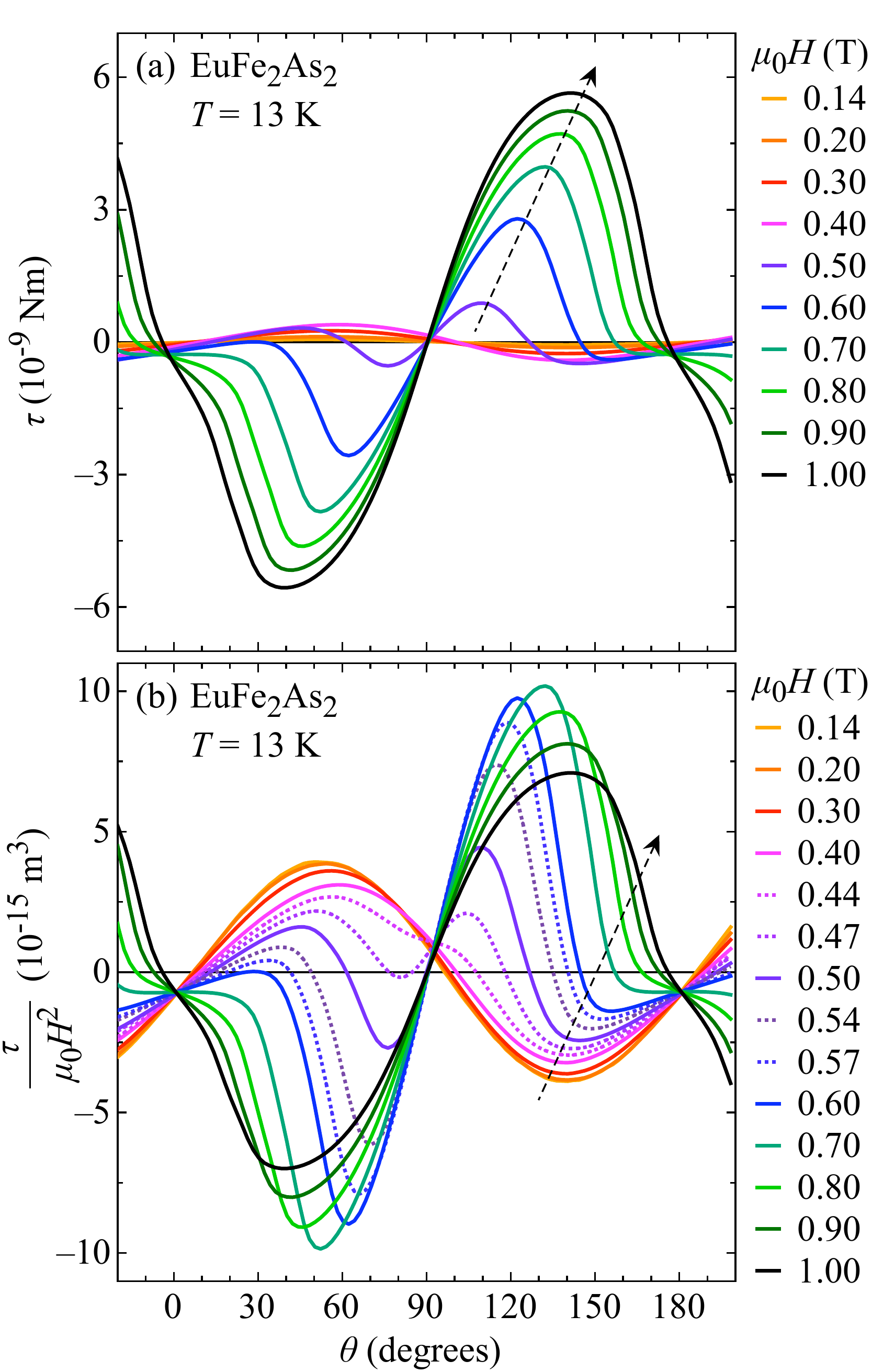}
\caption{(Color online) (a) Angular dependent magnetic torque $\tau$ 
of single-crystal EuFe$_2$As$_2$ at 13~K in various magnetic 
fields. For clarity not all measured data are shown. (b) Angular 
dependence of the quantity $\tau/(\mu_0H^2)$. The dashed arrows 
denote the direction of increasing magnetic field.}
\label{torque13Kund}
\end{figure}

 In low magnetic fields the Eu$^{2+}$ magnetic moments prefer to order 
antiferromagnetically in EuFe$_2$As$_2$. High magnetic fields reorient the magnetic moments, 
leading to  various magnetic field induced
phases. Magnetic torque allows to investigate multiple aspects
of magnetic order as a function of the magnetic field with respect to the principal axes. 
Whereas magnetization provides direct
information on the magnetic moment oriented along the
field, magnetic torque directly probes the anisotropy
of the susceptibility in magnetic systems. 
 
The angular dependence of the magnetic torque ${\tau}$ of single-crystal EuFe$_2$As$_2$
measured at 13~K in various magnetic fields is presented in Fig.~\ref{torque13Kund}a.
In Fig.~\ref{torque13Kund}b the same data are plotted in terms of $\tau / (\mu_0H^2$).
The torque data below 0.3~T are of sinusoidal
shape, following a simple angular dependence for a uniaxial antiferromagnet:\cite{Weyeneth}

\begin{figure}[b]
\centering
\includegraphics[width=1\linewidth]{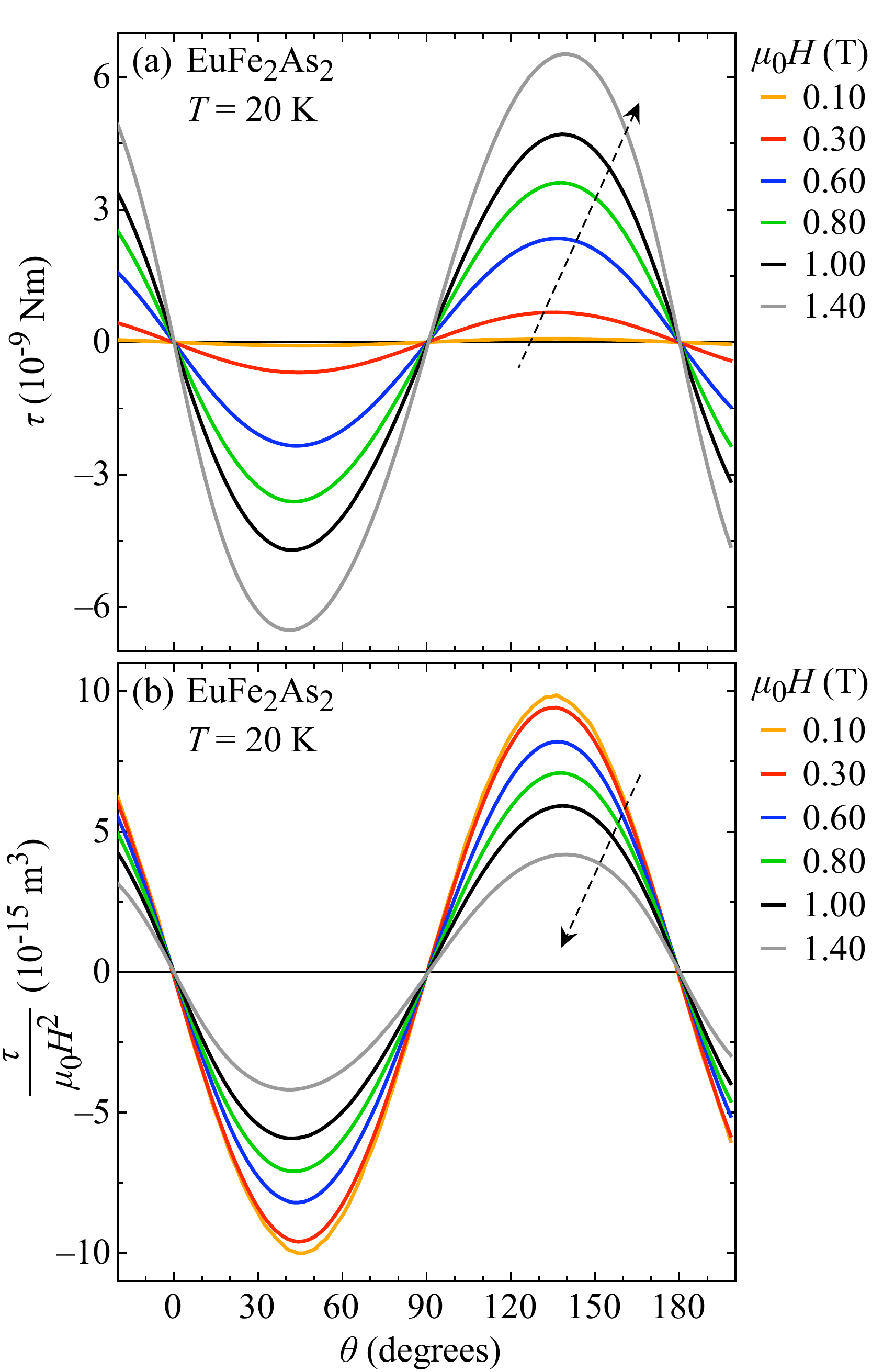}
\caption{(Color online) Magnetic torque ${\tau}$ (a) and the quantity $\tau / (\mu_0H^2$) (b)
as a function of the angle ${\theta}$ of single-crystal
EuFe$_2$As$_2$ in various magnetic fields at 20 K. 
The dashed arrows denote the direction of increasing magnetic field.}
\label{torque20Kund}
\end{figure}
  
\begin{figure*}[ht!]
\centering
\includegraphics[width=0.92\linewidth]{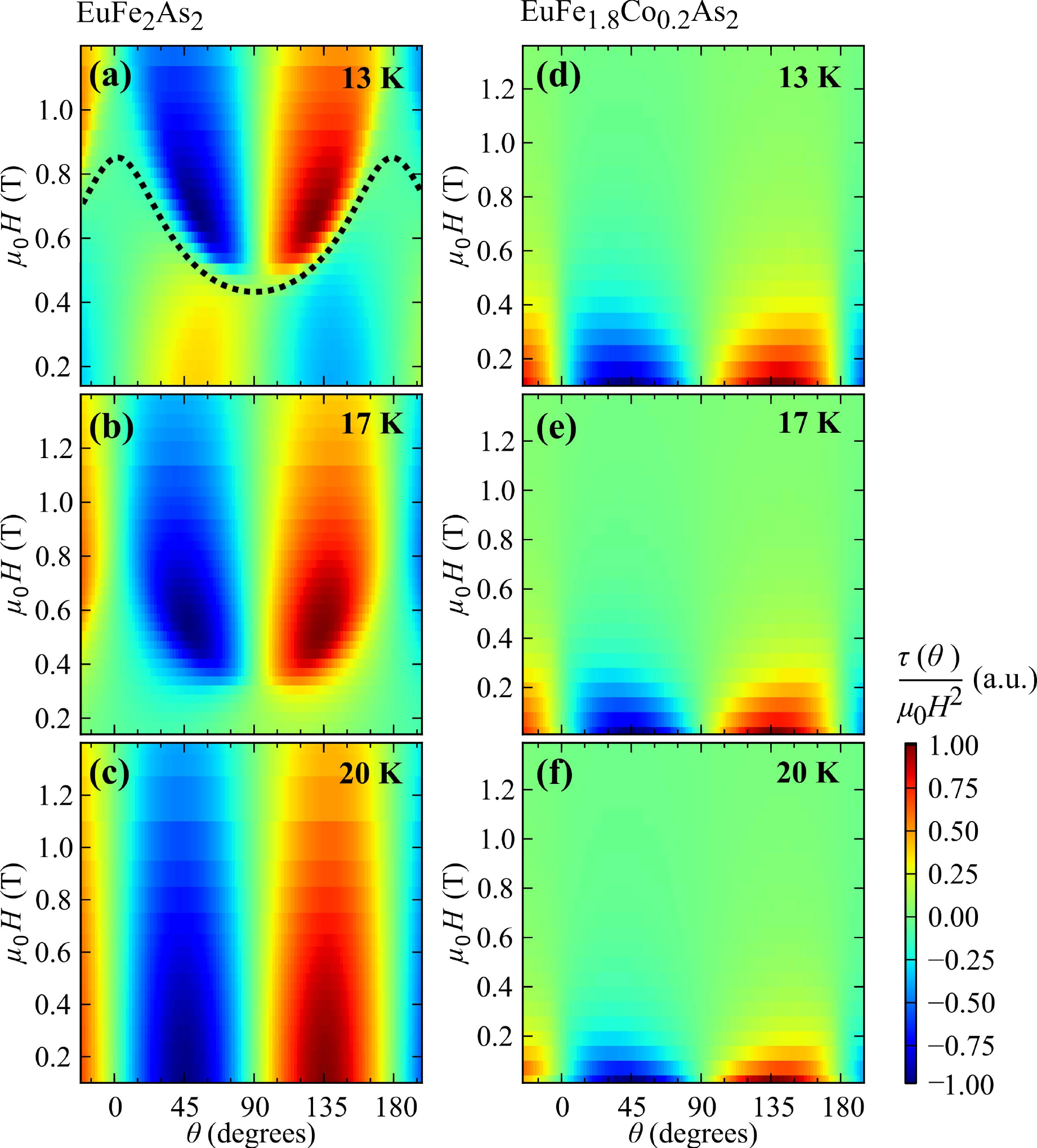}
\caption{(Color online) Color map of $\tau / (\mu_0H^2$) (in a.u.) for
EuFe$_2$As$_2$ and EuFe$_{1.8}$Co$_{0.2}$As$_{2}$ as a function of angle ${\theta}$ and field
$H$ for $T$ = 13 K, 17 K and 20 K. The dotted line in (a) is a fit
according to Eq.~(8). Panels (a), (b), and (c) are the
data for EuFe$_2$As$_2$ at 13 K, 17 K, and 20 K, respectively, whereas (d), (e) and (f)
are the data for EuFe$_{1.8}$Co$_{0.2}$As$_{2}$ at 13 K, 17 K, and 20 K,
respectively.}
\label{3Dtorque}
\end{figure*}

\begin{equation}
\tau(\theta) = -V\frac{\left(\chi_{\perp} - \chi_{\parallel}\right)}{2}\mu_0H^2\sin\left(2\theta\right). 
\end{equation}  
Here, $\theta$ denotes the angle between the field $H$ and the crystallographic
$c$-axis, $V$ is the volume of the sample, and ${\chi_{\perp}}$ and ${\chi_{\parallel}}$ 
are the magnetic susceptibilities for $H$ ${\perp}$ $c$ and for $H$ ${\parallel}$ $c$, respectively. 
Above 0.3~T the shape of the torque signal changes drastically (see Fig.~8).
For ${\theta}$  ${\simeq}$  90$^{\circ}$($H$ almost parallel to the $ab$-plane) an
additional torque signal appears,
with an opposite sign relative to the AFM torque. Upon increasing the magnetic field
this additional 
signal rises steeply and leads to
a sign change of the torque signal for all angles $\theta$. A similar
behavior was observed in RbVBr$_3$~\cite{Tanaka1992} and was interpreted as the
appearance of a weak field-induced magnetic moment. This additional contribution to
the torque signal observed here is
substantially larger than the AFM torque signal. This 
is consistent with the magnetization data (see Sec.~III(A)), from which the presence
of a
C-AFM phase was concluded above 0.3~T at 13~K.  The sign change of
the torque signal is in agreement with the sign change of the 
quantity $\chi_{\rm d} = \chi_{\perp} - \chi_{\parallel}$, 
which was interpreted as a signature of a transition to a C-AFM state of the
Eu$^{2+}$ magnetic moments.
It was shown previously \cite{Dengler10} that EuFe$_{2}$As$_{2}$ exhibits a weak
in-plane anisotropy. Since the in-plane anisotropy is much weaker than the
out-of-plane anisotropy, this system can be treated approximately as a uniaxial
anisotropic antiferromagnet. However, even a small in-plane anisotropy may lead to
discrepancies between experimental results and theoretical predictions for a
uniaxial anisotropic ferromagnet. Particularly, the torque signal of the AFM state
shown in 
Fig.~8a is shifted 
by ${\Delta}$${\theta}$ ${\sim}$ 10$^{\circ}$ with respect to one of the C-AFM state
(see Fig~8b). A similar phase shift 
${\Delta}$${\theta}$ was observed in
$\rm{\lambda}$ - (BETS)$_2$FeCl$_4$~\cite{Sasaki2001} and interpreted as a change of
the easy-axis.
However, here the phase shift
appears to indicate a crystallographic multi-domain state, due to a twinning of the
crystal in the AFM state. 

Figure~\ref{torque20Kund}a shows the
measured magnetic torque for the same EuFe$_2$As$_2$ single crystal at 20~K, where
according to the magnetization results the AFM regime has disappeared.
Consistently, no AFM torque signal is observed.
Instead, the magnetic torque amplitude increases with $H^2$ and saturates at higher 
$H$. Such a behavior is characteristic for a paramagnet. Consistently, the quantity
$\tau / (\mu_0H^2$) plotted in Fig.~9b decreases with increasing field. 
 
 In Fig.~\ref{3Dtorque} the scaled magnetic torque ${\tau}$/($\mu_0H^2$) for
EuFe$_2$As$_2$
and EuFe$_{1.8}$Co$_{0.2}$As$_{2}$ is shown in a
color map for the representative temperatures of 13 K, 17 K, and 20 K as a function
of angle ${\theta}$ and field $H$. 
Note that ${\tau}$/($\mu_0H^2$) is scaling according to 
the magnetic susceptibility. As seen in Fig.~\ref{3Dtorque}a the low field regime of
undoped EuFe$_2$As$_2$ at 13~K is dominated by the
AFM state, whereas for higher fields, the C-AFM state appears abruptly
along a clearly angular dependent boundary line (dotted line), demonstrating the
anisotropy of this magnetically ordered system. At 17 K (Fig.~\ref{3Dtorque}b) the AFM phase is not
present, consistent with the conclusions from the above susceptibility measurements. 
At 20~K (Fig.~\ref{3Dtorque}c) the signal is clearly
sinusoidal, consistent with FM behavior. 
In order to induce a canting of a planar antiferromagnetically ordered subsystem, 
the in-plane component of the magnetic field $H_\perp$ must surpass the in-plane 
magnetization $\mathcal{M}_\perp$ in one of the two magnetic sublattices:
\begin{equation}
H_\perp\geqslant A\cdot\mathcal{M}_\perp = A\cdot \sqrt{\mathcal{M}^2 - \mathcal{M}_\parallel^2}.
\end{equation}
Here, $\mathcal{M}$ is the saturation magnetization of the magnetic
sublattice, $\mathcal{M}_\parallel$ its out-of-plane component, and
$A$ is a constant. Taking into account
\begin{eqnarray}
H_{\perp} &=&H\sin(\theta),\\\nonumber
\mathcal{M}_{\parallel} &=& \frac{1}{2}{\chi}_{\parallel}H\cos(\theta),
\end{eqnarray}
 where ${\chi}_{\parallel}$ is the suceptibility of the total Eu$^{2+}$ magnetic sublattice, we
obtain for the boundary condition:
\begin{eqnarray}
H^2\sin^2(\theta)=A^2\left(\mathcal{M}^2-\frac{1}{4}\chi_\parallel^2H^2\cos^2(\theta)\right).
\end{eqnarray}
Solving this equality for $H$ yields the angle dependent canting field:
\begin{equation}\label{cantingtorque}
H_{\rm cant}(\theta)=\frac{A\cdot
\mathcal{M}}{\sqrt{\sin^2(\theta)+\frac{1}{4}\chi_{\parallel}^2A^2\cos^2(\theta})}.
\end{equation}
Interestingly, the resulting $H_{\rm cant}(\theta)$ is analog to the expression 
for the angular dependence of the upper critical field $H_{\rm c2}(\theta)$ in a type II superconductor.\cite{GinzburgLandau} Hence, Eq.~(\ref{cantingtorque}) can be simplified according to
\begin{equation}
H_{\rm cant}(\theta)=\frac{H_{\rm
cant,\perp}}{\sqrt{\sin^2(\theta)+\gamma_{\rm
cant}^{-2}\cos^2(\theta})},
\end{equation}
where $H_{\rm cant,\perp} = H_{\rm cant}(90^\circ)$ is the in-plane canting field, 
$\gamma_{\rm cant} = H_{\rm cant,\parallel}/H_{\rm cant,\perp}$ 
its anisotropy parameter, and $H_{\rm cant,\parallel} = H_{\rm cant}(0^\circ)$ the out-of-plane canting field.
This shape of the angular dependence of the transition between the AFM and C-AFM
phase in the ($H$, $\theta$) diagram is represented by the dashed line in Fig.~\ref{3Dtorque}a. 
It describes the experimental torque data rather well, with the 
parameters $H_{\rm cant},_{\perp}$(13 K) ${\simeq}$ 0.42(2) T and ${\gamma}$$_{\rm cant}$ ${\simeq}$ 2.0(2).  
This yields an estimate of the canting field parallel to the $c$-axis 
$H_{\rm cant},_{\parallel}$(13 K) ${\simeq}$ 0.84(6) T.

   The low field torque signal of EuFe$_{1.8}$Co$_{0.2}$As$_{2}$  
at 20 K (Fig.~10f) shows a shape typical for an anisotropic paramagnet. However, the
anisotropy of the
system is quite quickly suppressed with increasing magnetic field, which may
indicate a transformation of the paramagnetic state to a short range ordered state
at relatively low field. It might be caused
by large fluctuations of the magnetic moments in the vicinity of the transition from a
disordered PM state to an ordered one in EuFe$_{1.8}$Co$_{0.2}$As$_{2}$.
Furthermore, at low temperatures we do not observe any indication of a field induced
transition from 
the AFM to the C-AFM state (Fig.~10d-e). Therefore, we conclude that for
EuFe$_{1.8}$Co$_{0.2}$As$_{2}$
even at the lowest magnetic field a transition from a PM to a C-AFM state takes
place with decreasing temperature, in agreement with the above magnetization data.

\section{DISCUSSION}
\begin{figure*}[ht!]
\centering
\includegraphics[width=1\linewidth]{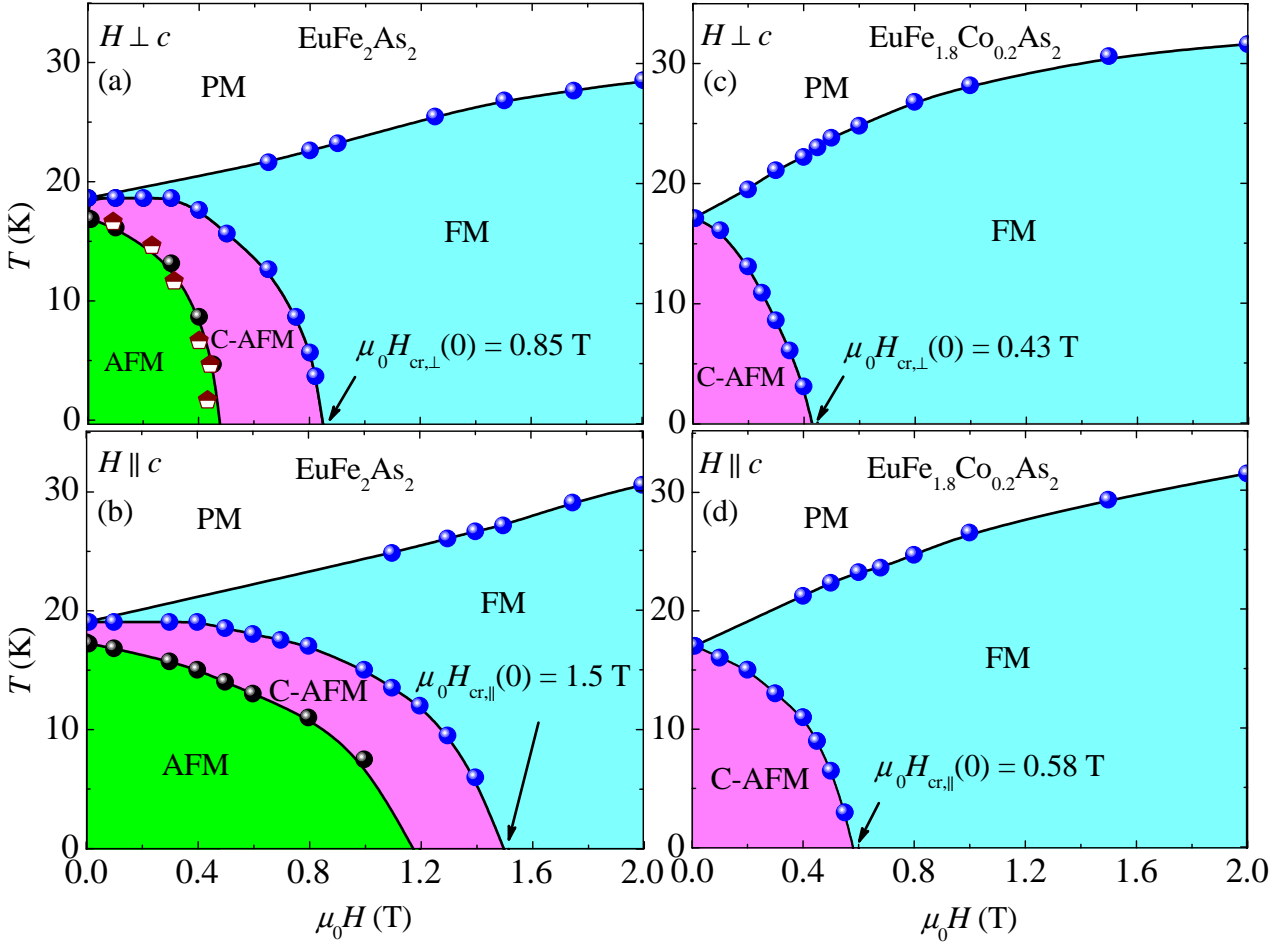}
\vspace{-0.2cm} \caption{ (Color online) Magnetic phase diagrams of single-crystal EuFe$_{2}$As$_{2}$ (a and b)  
and single-crystal EuFe$_{1.8}$Co$_{0.2}$As$_{2}$ (c and d) for \textit{H} ${\perp}$ $\textit{c}$ and for 
\textit{H} ${\parallel}$ $\textit{c}$. The various phases in the phase diagrams
are denoted as follows: paramagnetic (PM), antiferromagnetic (AFM), canted antiferromagnetic (C-AFM),
ferromagnetic (FM). The filled and open symbols are from the susceptibility and field dependent magnetization measurements, respectively. The solid lines are guides to the eyes.} \label{fig2}
\end{figure*}

 In Fig.~11 the results of the susceptibility, magnetization, and magnetic torque
experiments are summarized. They are discussed in terms of the phase diagram of the 
Eu$^{2+}$ magnetic sublattice of EuFe$_{2}$As$_{2}$ and EuFe$_{1.8}$Co$_{0.2}$As$_{2}$
for \textit{H} ${\perp}$ \textit{c} and \textit{H} ${\parallel}$ \textit{c}.

\subsection{EuFe$_{2}$As$_{2}$}
 For the parent compound EuFe$_{2}$As$_{2}$ four different magnetic phases were identified (see Fig.~11a and b):
a paramagnetic (PM), an antiferromagnetic (AFM), a canted antiferromagnetic (C-AFM),
and a ferromagnetic (FM) phase. The determination of the corresponding transition temperatures
and fields is described in Sec.~III.
The present experiments suggest a C-AFM order of the 
Eu$^{2+}$ spins in EuFe$_{2}$As$_{2}$ in the temperature range between 17 K and 19 K, while below 17 K 
an AFM structure is proposed. We suggest that at low temperatures
the system can be well described with a uniaxial model with easy plane and 
A-type AFM order.  
By applying a magnetic field within the AFM phase, a transition from AFM order via a canted configuration 
to a FM structure is observed. The observed $T_{\rm MM}$($H$) at which the metamagnetic (MM) transition
occurs (open symbols in Fig.~11a) is in agreement with the results obtained
from the susceptibility for the AFM to C-AFM transition (black filled symbols in Fig.~11a).
Thus, we propose that the MM transition corresponds to a spin-flop
transition from an AFM to a C-AFM state in EuFe$_{2}$As$_{2}$.
The critical magnetic field $H_{\rm cr}$($T$) at which the magnetic
moment in the Eu sublattice saturates was determined at different temperatures.
The values of $H_{\rm cr}$ extrapolated to zero temperature were found to be 
${\mu}$$_{0}$$H_{\rm cr},_{\rm \perp}$(0) ${\simeq}$ 0.85~T and 
${\mu}$$_{0}$$H_{\rm cr},_{\rm \parallel}$(0) ${\simeq}$ 1.5 T for \textit{H} ${\perp}$ $\textit{c}$ and 
\textit{H} ${\parallel}$ $\textit{c}$, respectively. By analyzing the shape of the angular dependence 
of $H_{\rm cr}$(${\theta}$) shown in Fig.~10a, we may conclude that the in-plane component of the 
magnetic field is responsible for the canting of the spins.
  
 The magnetic ordering of the Eu$^{2+}$ moments at low temperatures is consistent with the
magnetic structure established by neutron diffraction at 2.5 K.\cite{Xiao09} Note that in 
previous reports \cite{ZRen,SJiang} a possible C-AFM state in the 
temperature range 17 K ${\leq}$ $T$ ${\leq}$ 19 K was not discussed.  
To our knowledge no neutron data for the magnetic configuration of the 
Eu sublattice in this temperature range is available.

\subsection{EuFe$_{1.8}$Co$_{0.2}$As$_{2}$}
 The corresponding magnetic phase diagrams for Co-doped EuFe$_{1.8}$Co$_{0.2}$As$_{2}$  
are shown in Fig.~11c and d.
The magnetic ordering temperature of ${\simeq}$ 17 K is only about 2 K lower 
as compared to the parent compound. However, in the Co-doped EuFe$_{1.8}$Co$_{0.2}$As$_{2}$, no signatures
of a low-field and low-temperature AFM state of the Eu$^{2+}$ moments was found. 
Only a C-AFM phase (with a FM component in the $a$$b$-plane) is present at low fields and low temperatures. 
The ordering temperature $T_{\rm C-AFM}$ decreases with
increasing magnetic field, similar to the parent compound (see Fig.~11a and b). 
The critical magnetic field $H_{\rm cr}$ at which the Eu magnetic ordering
is saturated was determined for different temperatures,
and the extrapolated zero-temperature values were found to be: ${\mu}$$_{0}$$H_{\rm cr},_{\rm \perp}$(0) ${\simeq}$ 0.43~T 
and ${\mu}$$_{0}$$H_{\rm cr},_{\rm \parallel}$(0) ${\simeq}$ 0.58~T for \textit{H} ${\perp}$ $\textit{c}$ and \textit{H} ${\parallel}$ $\textit{c}$, respectively. These values of 
${\mu}$$_{0}$$H_{\rm cr}$ are much smaller than those obtained for the parent compound.
Moreover, the magnetic anisotropy 
$\gamma_{\rm cr}$ = $H_{\rm cr},_{\rm \parallel}$(0)/$H_{\rm cr},_{\rm \perp}$(0)  ${\simeq}$  1.35 of Co-doped EuFe$_{1.8}$Co$_{0.2}$As$_{2}$ is  
also smaller than $\gamma_{\rm cr}$ ${\simeq}$ 1.76 of the parent compound.  

 It was concluded from different experiments \cite{SJiang,Dengler10,Ying10,ZRen09,Guguchia} that 
there is a strong coupling between the localized Eu$^{2+}$ spins and the conduction electrons of the 
two-dimensional (2D) Fe$_{2}$As$_{2}$ layers. Recently, direct experimental evidence for a strong interlayer coupling was obtained by means of $^{75}$As NMR,\cite{Guguchia} revealing a magnetic exchange interaction between the localized Eu 4$f$ moments which is mediated by the itinerant Fe 3$d$ electrons. 
However, the direct interaction of the Eu moments and the magnetic moments 
in Fe sublattice cannot be neglected. Only a combination of both interactions can further 
elucidate the C-AFM ground state observed in the parent compound 
EuFe$_{2}$As$_{2}$ as well as in the Co-doped system EuFe$_{1.8}$Co$_{0.2}$As$_{2}$ (see Fig.~11). 

  Note that the present results for EuFe$_{1.8}$Co$_{0.2}$As$_{2}$, exhibiting a SDW ground state below 60 K,
\cite{Ying10} reveal a C-AFM structure of the Eu spins with a 
FM component in the $ab$-plane. This finding confirms previous assumptions that for materials in which 
the Fe ions are in the SDW ground state (such as EuFe$_{2}$As$_{2}$) 
the direction of the Eu magnetic moments is in the $ab$-plane.\cite{Xiao09,Nowik} 
On the other hand, in the case of non-magnetic Fe ground states, like in superconducting 
EuFe$_{2-x}$Co$_{x}$As$_{2}$ compounds, where the SDW magnetic state
is totally suppresed, the direction of the Eu magnetic moments 
is parallel to the $c$-axis.\cite{NowikF,Feng,NowikFel,NowikFelner}

\section{CONCLUSIONS}
 The magnetic properties of single crystals of EuFe$_{2}$As$_{2}$ and
EuFe$_{1.8}$Co$_{0.2}$As$_{2}$
were studied by means of susceptibility, magnetization, and magnetic torque investigations. 
The susceptibility and magnetization 
experiments performed for various temperatures and magnetic fields along 
the crystallographic axes provided information on the magnetic structure 
of the studied crystals. In addition, the evolution of the magnetic 
structure as a function of the tilting angle of the field and the 
crystallographic axes is studied by magnetic torque experiments.
The phase diagrams for the ordering of the Eu$^{2+}$ magnetic sublattice with 
respect to temperature, magnetic field and the angle between the magnetic field 
and the crystallographic $c$-axis in EuFe$_{2-x}$Co$_{x}$As$_{2}$ are determined and discussed.  
The present investigations reveal a complex and sophisticated interplay of magnetic phases in 
EuFe$_{2-x}$Co$_{x}$As$_{2}$. The magnetic ordering temperature
of the Eu$^{2+}$ moments remains nearly unchanged upon Co-doping. However, unlike the parent compound, in which
the Eu$^{2+}$ moments order antiferromagnetically at low temperatures,
the Co-doped system EuFe$_{1.8}$Co$_{0.2}$As$_{2}$ exhibits a C-AFM state with a FM component in the $ab$-plane. 
The magnetic anisotropy $\gamma_{\rm cr}$ becomes smaller as a result of Co-doping. This implies that the magnetic configuration of the Eu moments is strongly influenced by 
the magnetic moments of the Fe-sublattice, where superconductivity takes place for a certain range of Co-doping.  
A detailed knowledge of the interplay between the Eu$^{2+}$ moments
and magnetism of the Fe sublattice is important to understand the role of magnetism of the
localized Eu$^{2+}$ moments for the occurence of supercondcutivity in EuFe$_{2-x}$Co$_{x}$As$_{2}$.\\

\section{Acknowledgments}
This work was supported by the Swiss National Science Foundation, the
SCOPES grant No. IZ73Z0${\_}$128242, the NCCR Project MaNEP, the EU Project
CoMePhS, and the Georgian National Science Foundation grant
GNSF/ST08/4-416.





\begin{thebibliography}{99}
\bibitem{Kamihara08} Y.~Kamihara, T.~Watanabe, M.~Hirano, and H.~Hosono, J. Am.
Chem. Soc. \textbf{130}, 3296 (2008).

\bibitem{Chen08} X.H.~Chen, T.~Wu, G.~Wu, R.H.~Liu, H.~Chen, and D.F.~Fang,
Nature (London) \textbf{453}, 761 (2008).






\bibitem{Rotter} M.~Rotter, M.~Tegel, and D.~Johrendt, Phys. Rev. Lett. \textbf{101}, 107006 (2008). 

\bibitem{Hsu} F.-C.~Hsu, J.-Y.~Luo, K.-W.~Yeh, T.-K.~Chen, T.-W.~Huang, P. M.~Wu, Y.-C.~Lee, Y.-L.~Huang, Y.-Y. ~Chu, D.-C.~Yan, and M.-K.~Wu, Proc. Natl. Acad. Sci. USA \textbf{105}, 14263 (2008).


\bibitem{Xiao09} Y.~Xiao, Y.~Su, M.~Meven, R.~Mittal, C.M.N.~Kumar, 
T.~Chatterji, S.~Price, J.~Persson, N.~Kumar, S.K.~Dhar, 
A.~Thamizhavel, and Th.~Brueckel, Phys. Rev. B  \textbf{80}, 174424  (2009).

\bibitem{Torikachvili} M.S.~Torikachvili, S.L.~Bud'ko, N.~Ni, and P.C.~Canfield, Phys.
Rev. Lett. \textbf{101}, 057006 (2008).

\bibitem{Sun} Liling Sun, Jing Guo, Genfu Chen, Xianhui Chen, Xiaoli Dong, Wei Lu, Chao Zhang, Zheng Jiang, Yang Zou, Suo Zhang, Yuying Huang, Qi Wu, Xi Dai, Yuanchun Li, Jing Liu, and Zhongxian Zhao,
Phys. Rev. B  \textbf{82}, 134509  (2010).

\bibitem{Lee} H.~Lee, E.~Park, T.~Park, V.A.~Sidorov, F.~Ronning, E.D.~Bauer,
and J.D.~Thompson, Phys. Rev. B \textbf{80}, 024519 (2009).

\bibitem{Alireza} P.L.~Alireza, Y.T.C.~Ko, J.~Gillett, G.G.~Lonzarich, and 
S.E.~Sebastian, J. Phys.: Condens. Matter \textbf{21}, 012208 (2009).

\bibitem{Igawa} K.~Igawa, H.~Okada, H.~Takahahsi, S.~Matsuishi, Y.~Kamihara,
M.~Hirano, H.~Hosono, K.~Matsubayashi, and Y.~Uwatoko, J. Phys. Soc. Jpn. \textbf{78}, 025001 (2009).

\bibitem{Fukazawa} H.~Fukazawa, N.~Takeshita, T.~Yamazaki, K.~Kondo, K.~
Hirayama, Y.~Kohori, K.~Miyazawa, H.~Kito, H.~Eisaki, and A.~Iyo,
J. Phys. Soc. Jpn. \textbf{77}, 105004 (2008).

\bibitem{Duncan} W.J.~Duncan, O.P.~Welzel, C.~Harrison, X.F.~Wang, 
X.H.~Chen, F.M.~Grosche, and P.G.~Niklowite, J. Phys.: Condens.
Matter \textbf{22}, 052201 (2010).

\bibitem{Mani} A.~Mani, N.~Ghosh, S.~Paulraj, A.~Bharathi, and C.S.~Sundar,
Europhys. Lett. \textbf{87}, 17004 (2009).


\bibitem{Terashima} T.~Terashima, M.~Kimata, H.~Satsukawa, A.~Harada,
K.~Hazama, S.~Uji, H.S.~Suzuki, T.~Matsumoto, and K.~Murata,
J. Phys. Soc. Jpn. \textbf{78}, 083701 (2009).

\bibitem{Miclea} C.F.~Miclea, M.~Nicklas, H.S.~Jeevan, D.~Kasinathan, Z.~Hossain,
H.~Rosner, P.~Gegenwart, C.~Geibel, and F.~Steglich, Phys.
Rev. B \textbf{79}, 212509 (2009).

\bibitem{RenLu08} Z.A.~Ren, W.~Lu, J.~Yang, W.~Yi, X.L.~Shen, Z.C.~Li, 
G.C.~Che, X.L.~Dong, L.L.~Sun, F.~Zhou, and Z.X.~Zhao,
Chin. Phys. Lett.  \textbf{25}, 2215 (2008).

\bibitem{Matsuishi08} S.~Matsuishi, Y.~Inoue, T.~Nomura, M.~Hirano, and 
H.~Hosono, J. Phys. Soc. Jpn.  \textbf{77}, 113709 (2008).

\bibitem{Zhao08} J.~Zhao, Q.~Huang, C.~de la Cruz, S.~Li, J.W.~Lynn, 
Y.~Chen, M.A.~Green, G.F.~Chen, G.~Li, Z.~Li, J.L.~Luo,
N.L.~Wang, and P.~Dai, Nature Materials  \textbf{7}, 953 (2008).

\bibitem{Raffius93} H.~Raffius, M.~M\"{o}rsen, B.D.~Mosel, W.~M\"{u}ller-Warmuth,
W.~Jeitschko, L.~Terb\"{u}chte, and T.~Vomhof, J. Phys.
Chem. Solids  \textbf{54}, 135 (1993).

\bibitem{ZRen} Z.~Ren, Z.W.~Zhu, S.~Jiang, X.F.~Xu, Q.~Tao, C.~Wang, C.M.~Feng, G.H.~Cao, and Z.A.~Xu, Phys. Rev. B \textbf{78}, 052501 (2008).

\bibitem{SJiang} S.~Jiang, Y.K.~Luo, Z.~Ren, Z.W.~Zhu, C.~Wang, X.F.~Xu, 
Q.~Tao, G.H.~Cao, and Z.A.~Xu, New J. Phys. \textbf{11}, 025007 (2009).

\bibitem{Blundell} S. Blundell, \textit{Magnetism in Condensed Matter} (New York: Oxford University Press, 2006).

\bibitem{Xiao} Y.~Xiao, Y.~Su, W.~Schmidt, K.~Schmalzl, C.M.N.~Kumar, S.~Price, 
T.~Chatterji, R.~Mittal, L.J.~Chang, S.~Nandi, N.~Kumar, S.K.~Dhar, 
A.~Thamizhavel, and Th.~Brueckel, 
Phys. Rev. B \textbf{81}, 220406 (2010). 

\bibitem{ShuaiJiang} Shuai Jiang, Hui Xing, Guofang Xuan, Zhi Ren, Cao Wang, Zhu-an Xu, and Guanghan Cao,
Phys. Rev. B \textbf{80}, 184514 (2009).

\bibitem{He08} Y.~He, T.~Wu, G.~Wu, Q.J.~Zheng, Y.Z.~Liu, H.~Chen, J.J.~Ying,
R.H.~Liu, X.F.~Wang, Y.L.~Xie, Y.J.~Yan, J.K.~Dong, S.Y.~Li, and
X.H.~Chen, J. Phys.: Condens. Matter  \textbf{22}, 235701 (2010).

\bibitem{Jeevan} H.S.~Jeevan, Deepa~Kasinathan, H.~Rosner, and P.~Gegenwart, 
Phys. Rev. B  \textbf{83}, 054511  (2011).

\bibitem{ZRen09} Zhi~Ren, Xiao~Lin, Qian~Tao, Shuai~Jiang, Zengwei~Zhu, Cao~Wang, Guanghan~Cao, and Zhu'an Xu, Phys. Rev. B  \textbf{79}, 094426 (2009). 

\bibitem{LiLuo08} L.J.~Li, Y.K.~Luo, Q.B.~Wang, H.~Chen, Z.~Ren, Q.~Tao, 
Y.K.~Li, X.~Lin, M.~He, Z.W.~Zhu, G.H.~Cao, and Z.A.~Xu, 
New J. Phys.  \textbf{11}, 025008 (2009).


\bibitem{Dengler10} E.~Dengler, J.~Deisenhofer, H.A.~Krug von Nidda, Seunghyun~Khim, J.S.~Kim, Kee Hoon Kim, 
F.~Casper, C.~Felser, and A.~Loidl, Phys. Rev. B  \textbf{81}, 024406  (2010).  

\bibitem{Ying10} J.J.~Ying, T.~Wu, Q.J.~Zheng, Y.~He, G.~Wu, Q.J.~Li, Y.J.~Yan, Y.L.~Xie, R.H.~Liu, X.F.~Wang, 
and X.H.~Chen, Phys. Rev. B  \textbf{81}, 052503 (2010). 





\bibitem{Guguchia} Z.~Guguchia, J.~Roos, A.~Shengelaya, S.~Katrych, Z.~Bukowski, S.~Weyeneth,
F.~Mur\'{a}nyi, S.~Str\"{a}ssle, A.~Maisuradze, J.~Karpinski, and H.~Keller, 
Phys. Rev. B  \textbf{83}, 144516 (2011).

\bibitem{Nowik} I. Nowik and I. Felner, \textit{Hyperfine Interact.} \textbf{28}, 959 (1986).

\bibitem{NowikF} I. Nowik and I. Felner,  \textit{Physica} C \textbf{469}, 485 (2009).

\bibitem{Feng} C. Feng, Z. Ren, S. Xu, S. Jiang, Z. Xu, G. Cao, I. Nowik, I. Felner, K. Matsubayashi, and 
Y. Uwatoko, Phys. Rev. B \textbf{82}, 094426 (2010).

\bibitem{NowikFel} I. Nowik, I. Felner, Z. Ren, G.H. Cao, and Z.A. Xu, J. Phys.: Condens. Matter \textbf{23}, 065701 (2011).

\bibitem{NowikFelner} I. Nowik, I. Felner, Z. Ren, G.H. Cao, and Z.A. Xu, New J. Phys. \textbf{13}, 023033 (2011).

\bibitem{Kohout2007} S.~Kohout, J.~Roos, and H.~Keller, Rev.~Sci.~Instrum. \textbf{78}, 013903 (2007).




\bibitem{Luo} J.L.~Luo, N.L.~Wang, G.T.~Liu, D.~Wu, X.N.~Jing, F.~Hu, and T.~Xiang, Phys. Rev. Lett. \textbf{93}, 187203 (2004). 

\bibitem{Perry} R.S.~Perry, L.M.~Galvin, S.A.~Grigera, L.~Capogna, A.J.~Schofield, A.P.~Mackenzie, M.~Chiao, 
S.R.~Julian, S.I.~Ikeda, S.~Nakatsuji, Y.~Maeno, and C.~Pfleiderer, Phys. Rev. Lett. \textbf{86}, 2661 (2001).

\bibitem{Kimura} T.~Kimura and Y.~Tokura, Ann. Rev. Mater. Sci. \textbf{30}, 451 (2000).

\bibitem{Weyeneth}  S.~Weyeneth, P.J.W. Moll, R.~Puzniak, K.~Ninios, F.F.~Balakirev, R.D.~McDonald, H.B.~Chan, 
N.D.~Zhigadlo, S.~Katrych, Z.~Bukowski, J.~Karpinski, H.~Keller, B.~Batlogg, and L.~Balicas, 
Phys. Rev. B \textbf{83}, 134503 (2011) 

\bibitem{Tanaka1992} H.~Tanaka, T.~Kato, K.~Iio, and K.~Nagata, J. Phys. Soc. Jap. \textbf{61}, 3292 (1992).

\bibitem{Sasaki2001} T.~Sasaki, H.~Uozaki, S.~Endo, and N.~Toyota, Synth. Metals \textbf{120}, 759 (2001).

\bibitem{GinzburgLandau} V.~L.~Ginzburg and L.~D.~Landau, Zh. Eksp. Teor. Fiz. {\bf 20}, 1064 (1950).

\end{thebibliography}
\end{document}